\documentclass[letterpaper, journal]{IEEEtran}

\usepackage{amsmath,amsthm,graphicx,amssymb,fullpage,wrapfig}

\usepackage{setspace}
\usepackage{cite}

\usepackage{subfig}

\usepackage{array}
\usepackage{amsmath,amsthm,amssymb,algorithmic,algorithm}
\usepackage[margin=1in]{geometry}
\usepackage{setspace}
\usepackage{color}
\usepackage{soul}
\usepackage{epstopdf}

\usepackage{adjustbox}
\usepackage{bm}

\theoremstyle{remark}

\newcommand{\ben}{\begin{eqnarray}}

\newcommand{\een}{\end{eqnarray}}

\newcommand{\ab}{{a}}
\newcommand{\bb}{{b}}

\title{VESPA: VIPT Enhancements for \\Superpage Accesses}
\author{Mayank Parasar$^\dag$, Abhishek Bhattacharjee$^\ddag$ and Tushar Krishna$^\dag$\\
\small $^\dag$School of ECE, Georgia Institute of Technology \hspace{4mm} $^\ddag$Dept. of CS, Rutgers University\\
\small mparasar3@gatech.edu \hspace{2mm} abhib@cs.rutgers.edu \hspace{2mm} tushar@ece.gatech.edu  
}

\begin{document}
%
\maketitle
\begin{abstract}
L1 caches are critical to the performance of modern computer
systems. Their design involves a delicate balance between fast
lookups, high hit rates, low access energy, and simplicity of
implementation. Unfortunately, constraints imposed by virtual memory
make it difficult to satisfy all these attributes today. Specifically,
the modern staple of supporting virtual-indexing and physical-tagging
(VIPT) for parallel TLB-L1 lookups means that L1 caches are usually
grown with greater associativity rather than sets.  This compromises
performance -- by degrading access times without significantly
boosting hit rates -- and increases access energy.

We propose {\sf \bf VIPT Enhancements for SuperPage Accesses} or
 {\sf \bf VESPA} in response. {\sf VESPA} side-steps the traditional
 problems of VIPT by leveraging the increasing ubiquity of
 superpages\footnote{By superpages (also called huge or large pages)
 we refer to any page sizes supported by the architecture bigger than
 baseline page size.}; since superpages have more page offset bits,
 they can accommodate L1 cache organizations with more sets than
 baseline pages can. {\sf VESPA} dynamically ada\-pts to any OS
 distribution of page sizes to operate L1 caches with good access
 times, hit rates, and energy, for both single- and multi-threaded
 workloads. Since the hardware changes are modest, and there are no OS
 or application changes, {\sf VESPA} is readily-implementable.
\end{abstract}

\section{Introduction}\label{sec:introduction}
{

As processors keep integrating more CPUs, the cache subsystem
continues to critically affect system performance and energy. Modern
processors employ several levels of caches, but the L1 cache remains
important.  L1 caches are system-critical because they service {\it
  every single} memory reference, whether it is due to a CPU lookup or
a coherence lookup.  L1 caches must balance the following design
objectives:

\vspace{2mm} \noindent \textcircled{1} {\bf Good performance:} L1
caches must achieve high hit rates and low access times. This requires
balancing the number of sets and set-associativity; for example,
higher associativity can increase hit rates, but worsen access times.

\vspace{2mm} \noindent \textcircled{2} {\bf Energy efficiency:} Cache hit,
miss, and management (e.g., insertion, coherence, etc.) energy must be
minimized. These too require a balance; for example, higher
set-associa\-tivity may reduce cache misses and subsequent energy-hun\-gry
probes of larger L2 caches and LLCs. But higher set-assoc\-iativity also
magnifies L1 lookup energy.

\vspace{2mm} \noindent \textcircled{3} {\bf Simple implementation:} L1
caches are on the pipeline's timing-critical path. To meet timing
constraints, they should not have complex implementations.

\vspace{2mm} Balancing these design goals is challenging for many
reasons \cite{L1, L1_1, L1_2, L1_3}. A particularly important one is
the L1 cache's interaction with the virtual memory subsystem. Virtual
memory automates memory and storage management, but requires
virtual-to-physical address translation. CPUs accelerate address
translation with hardware Translation Lookaside Buffers
(TLBs). Ideally, TLBs should be large and cache as many translations
as possible. Unfortunately, this presents a problem. L1 caches are
physically-addressed, and hence require address translations via the
TLB {\it prior} to a cache lookup.  Large and slow TLBs can delay L1
cache lookup time.

\begin{figure}
\setlength{\abovecaptionskip}{0pt}
\setlength{\belowcaptionskip}{0pt}
\begin{center}
    \begin{minipage}{0.48\textwidth}
\includegraphics[width=1\linewidth]{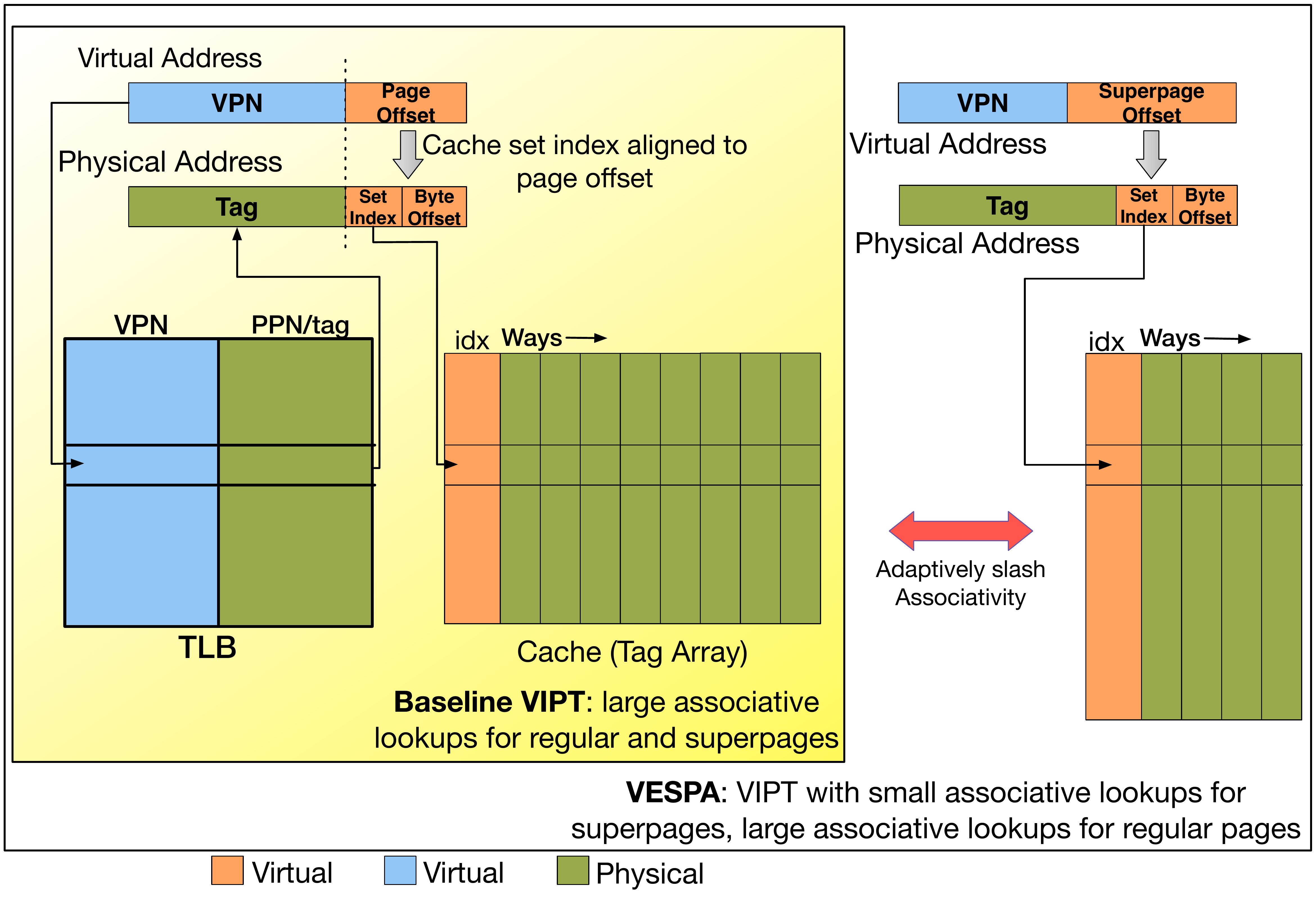}
\caption{\small \bf Overview of  VESPA 
 compared to traditional VIPT caches. VESPA 
dynamically changes its associativity for superpages.}
\vspace{-4.5mm}
\label{fig:high_level_arch}
\end{minipage}
\end{center}
\end{figure}

Architects have historically used virtual-indexing and
phy\-sical-tagging (VIPT) to circumvent this problem. VIPT L1 caches are
looked up in parallel with -- rather that serially after -- TLBs. The
basic idea is that TLB lookup proceeds in parallel with L1 cache set
selection, and must be completed by the time the 
L1 cache begins checking the
tags of the ways in the set. On one hand, this allows architects to
grow TLBs bigger. On the other hand, VIPT presents challenges in
balancing \textcircled{\small 1}-\textcircled{\small 3}. The key
problem is this -- in VIPT designs, the bits used to select the cache
set must reside entirely within the page offset of the requested
virtual address, as shown in Figure~\ref{fig:high_level_arch}. 
This is because the page-offset bits remain the same in both the 
virtual and the physical address.
This in
turn means that the VIPT cache organization rests at the mercy of the
operating system's (OS's) page size.

Consider, without loss of generality, x86 systems, where the smallest
or {\it base} page size is 4KB. The 12 least significant bits of the
virtual address correspond to the page offset.  VIPT L1 caches with
64-byte lines need 6 bits for the cache block offset. Unfortunately,
this only leaves 6 bits for the set index, meaning that the L1 cache
can support a maximum of 64 sets. Consequently, VIPT L1 cache sizes
can only be increased through higher set-associativity.  This can
compromise performance (\textcircled{1}) and energy (\textcircled{2})
significantly, especially as cache sizes increase. We identify and
quantify the extent of this problem this work via a state-space
exploration of cache microarchitectures with varying sizes and
set-associativity (Section~\ref{sec:cacti_acctime_energy}) and find
that it is pernicious.

Past work has tackled the limitations of VIPT with techniques like
virtually-indexed, virtually-tagged (VIVT) caches \cite{SynonymFilter,
  Segmented_addr}, reducing L1 cache associativity by pushing part of
the index into physical page number \cite{VirCache}, opportunistic
virtual caching \cite{ovc} and similar designs which argue that VIPT
can be simplified by observing that page synonyms\footnote{Synonyms
  are scenarios where multiple virtual addresses map to the same
  physical address.} are rare \cite{VirCache, two_level, in_cache,
  energy_eff, slb}.  While effective, these techniques are complex,
falling short on \textcircled{3}. Unsurprisingly, VIPT therefore
remains the standard in modern systems.

We propose \textsf{\textbf{VIPT Enhancements for Super Page Accesses}}
or \textsf{\textbf{VESPA}} to free L1 caches from the shackles of VIPT
design. {\sf VESPA} hinges on the following observation. VIPT L1
caches were originally designed when OSes supported one page
size. However, systems today support and use not only these
traditional or {\it base} pages, but also several larger pages (e.g.,
2MB and 1GB on x86 systems) called {\bf superpages} \cite{THP_redhat,
  THP}. Superpages have historically been used to reduce TLB misses.
In this work, we go beyond, and leverage the prevalence of superpages
to realize VIPT caches that can achieve \textcircled{\small 1},
\textcircled{\small 2}, and \textcircled{\small 3}.

Superpages have wider page offsets and can therefore support more
index bits for VIPT; this in turn means that the same sized cache can
be realized with more sets and lower associativity.  {\sf VESPA}
harnesses this property to dynamically reduce associativity within a
set for superpages, as shown in
Figure~\ref{fig:high_level_arch}. Consequently, {\sf VESPA} improves
performance (\textcircled{1}) and energy efficiency (\textcircled{2}).
Moreover, it leverages the current TLB-L1 interface, and only modestly
changes the L1 microarchitecture, making it easy to implement
(\textcircled{3}). In other words, {\sf VESPA} {\it improves}, but
still conceptually uses the concept of VIPT, and satisfies three types
of lookups:

\vspace{2mm}\noindent{\textcircled{\small a} \bf CPU lookups for data
  in a superpage:} {\sf VESPA} checks fewer L1 cache ways than
traditional VIPT caches, reducing hit time and saving energy.

\vspace{2mm}\noindent{\textcircled{\small b} \bf CPU lookups for data
  in a base page:} {\sf VESPA} checks the same number of L1 cache
ways as traditional VIPT caches, and achieves the same performance and
energy.

\vspace{2mm}\noindent{\textcircled{\small c} \bf Coherence lookups:}
Though coherence lookups at L1 use physical addresses and 
hence do not need
to look up the TLB, they 
needlessly have look up the high number of ways in the L1s because the L1s adhere to VIPT constraints.
{\sf VESPA} solves
this problem, allowing {\it all} coherence lookups, whether they are
to addresses in superpages or base pages, to check fewer L1 cache
ways. Consequently, coherence energy is reduced substantially.

\vspace{2mm}
\noindent Overall, our contributions are:

\begin{itemize}[itemsep=0in]
\vspace{1.5mm}\item We study cache lookup time and energy in detail
with Cacti 6.5~\cite{cacti6.5}, motivating the need to design caches
with lower associativities than those supported by VIPT.

\vspace{1.5mm}\item We perform a real-system characterization study on
the prevalence of superpages in modern systems. We find that current
Linux/x86 systems create ample superpages for {\sf VESPA} to be
effective.

\vspace{1.5mm}\item We showcase {\sf VESPA's} benefits for CPU memory
loo\-kups. On a wide suite of single- and multi-threaded workloads
running on real systems and full-system simulators, {\sf VESPA}
demonstrates the following improvements in performance and energy over
a baseline VIPT 32-64kB L1 cache: 5-12\% in AMAT, 9-18\% in dynamic
energy, 95-98\% in leakage energy.  These go up to 19\%, 23\%, and
99\% in larger forward-looking 128kB L1s.  Further, we show how these
benefits change as the prevalence of superpages varies.

\vspace{1.5mm}\item We quantify {\sf VESPA's} energy benefits for L1
coherence lookup, showing significant benefits (e.g., 43\% L1 energy
savings on a 64-core system).

\end{itemize}
\vspace{1.5mm}

To the best of our knowledge, this is the first work to optimize L1
caches for superpages.  {\sf VESPA} improves \textcircled{\small 1},
\textcircled{\small 2}, and \textcircled{\small 3} comprehensively.

Importantly, {\sf VESPA} improves current systems, but is likely to be
even more crucial as future systems seek ways to increase L1 cache
sizes~\cite{excavator} while meeting VIPT constraints.

\section{Motivation and Background}
\label{sec:motivation}

\begin{figure*}[tb]
\setlength{\abovecaptionskip}{0pt}
\setlength{\belowcaptionskip}{0pt}
\centering
\subfloat[\small{\bf Cache Access Latency}]
{
\includegraphics[width=0.34\linewidth]{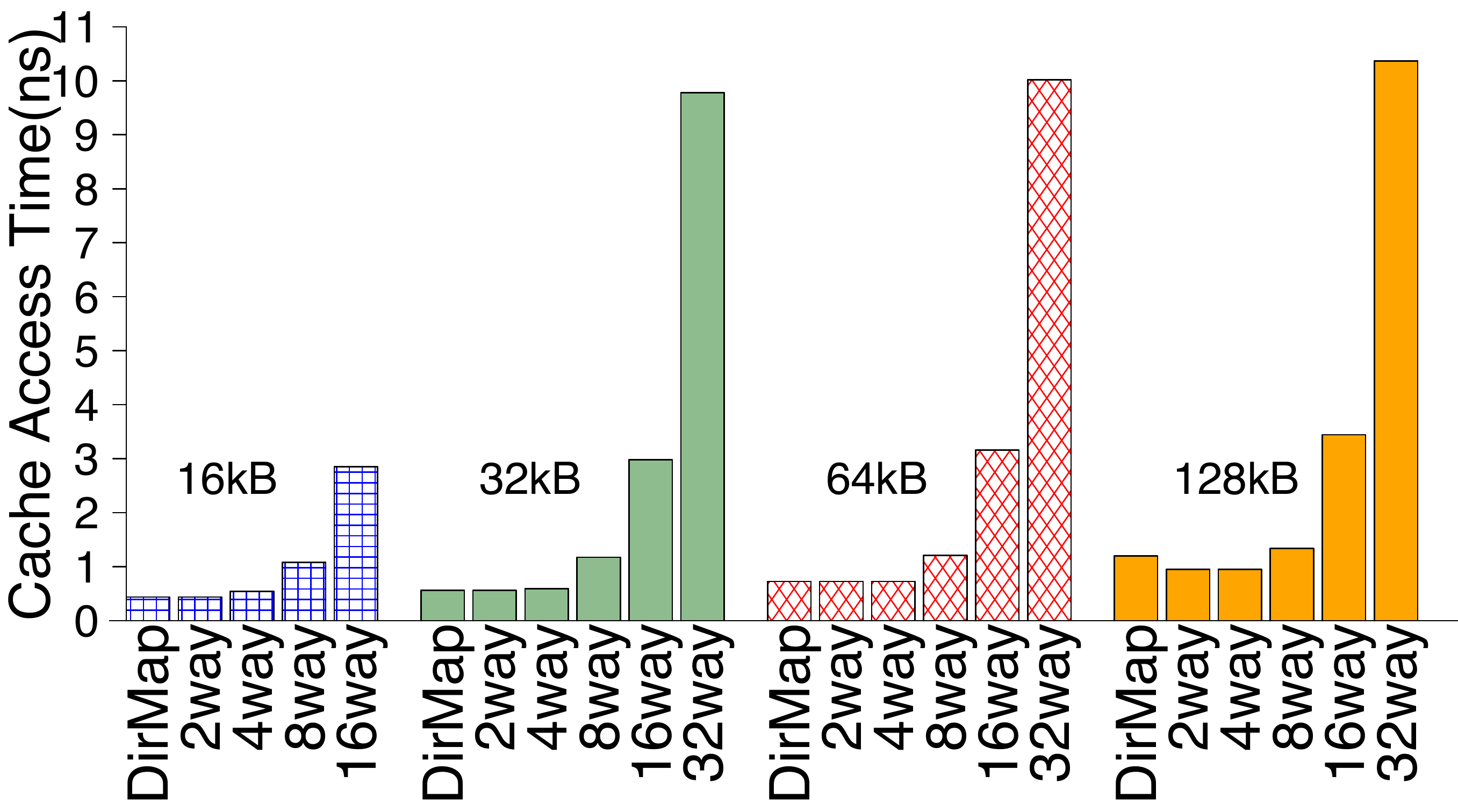}
\label{fig:acc_time}
}
\subfloat[\small{\bf  Cache access energy (dynamic + leakage)}]
{
\includegraphics[width=0.34\linewidth]{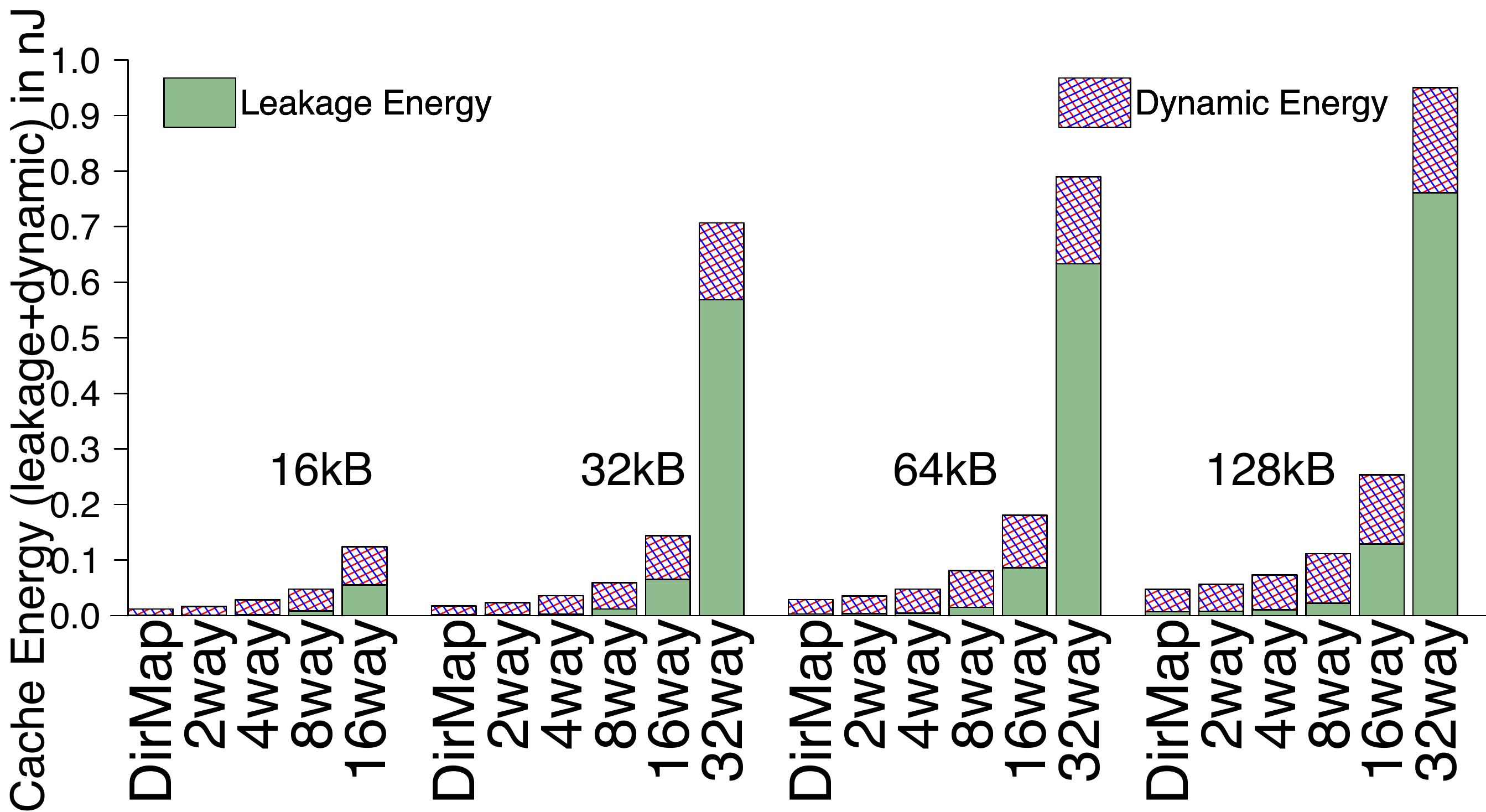}
\label{fig:acc_energy}
}
\subfloat[\small{\bf Average Miss-per-kilo-instructions (MPKI)}]
{
\includegraphics[width=0.34\linewidth]{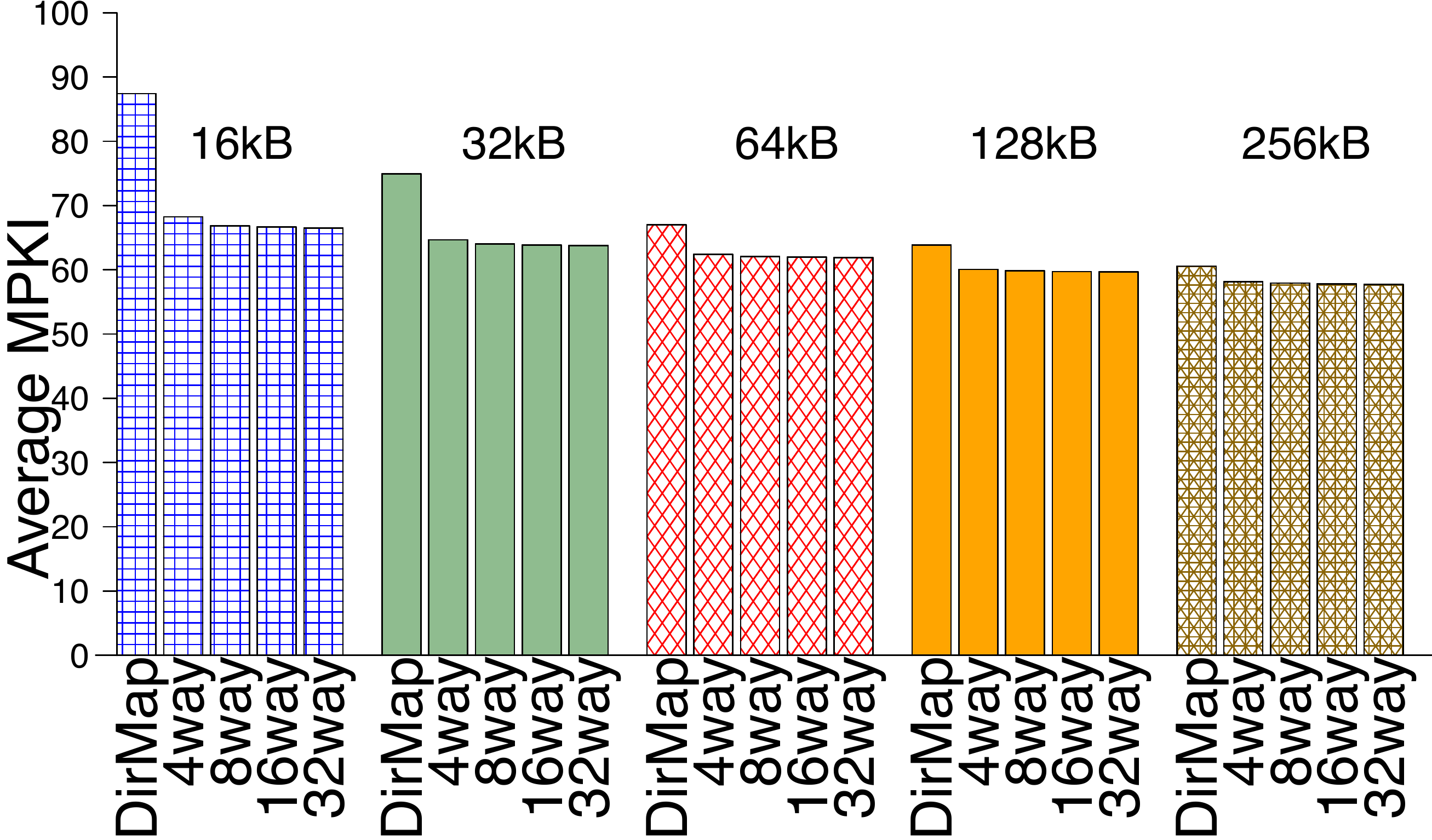}
\label{fig:mpki}
}
\caption{\small \sf \bf Effect of cache latency, energy, and MPKI as a function of associativity for different cache sizes.}
\label{fig:fragmentation_sweep}

\end{figure*}

\subsection{TLBs and VIPT Caches}
\label{sec:vipt}
When CPUs access memory, they typically do so using program-visible
virtual memory addresses (VA). These must be converted into physical
memory addresses (PA). The OS maintains page tables to to record
virtual-to-physical address mappings in the unit of pages. Since page
tables are software data structures, they reside in the on-chip caches
and system memory. CPUs cache the most frequently-used translations
from the page table in private TLBs. Therefore when a CPU accesses
memory using a VA, the TLB is probed to determine the PA. A TLB miss
invokes a hardware page table walker, which looks up the page
table. In general, TLBs interface with L1 caches in 3 ways:

\vspace{2mm}\noindent{\textcircled{\small a} \bf Physically-indexed
  physically-tagged (PIPT):} L1 accesses commence only after TLBs are
looked up to determine the physical address. Though simple to
implement, serial TLB-L1 lookups are slow and rarely used.

\vspace{2mm}\noindent{\textcircled{\small b} \bf Virtually-indexed
  virtually-tagged (VIVT):} L1 accesses do not need a prior TLB
lookup, since they operate purely on virtual addresses. Unfortunately,
VIVT caches are complex, requiring special support for virtual page
synonyms and cache coherence (which typically uses physical
addresses). We discuss this design further in Section~\ref{sec:related}.

\vspace{2mm}\noindent{\textcircled{\small c} \bf Virtually-indexed
  physically-tagged (VIPT):} By far the most popular choice, VIPT
caches strike a compromise between PIPT and VIVT. With VIPT, the VA is
used to index the L1 while the TLB is probed in parallel for the PA;
TLB access latency is hidden. However, to support physically-addressed
L1s, the cache index bits must fall {\it within} the page offset bits,
as shown in Figure \ref{fig:high_level_arch}.  This restricts the
number of cache sets. Hence, the cache can grow only by increasing
associativity (i.e., adding more ways).  {\it This is reflective in
  all modern systems. Intel Skylake~\cite{Intel_skylake} uses an 8-way
  cache to implement its 32kB L1 and AMD's Jaguar uses a 16-way cache
  (8 banks of 2-ways each) to implement its 64kB L1 cache, since the
  base page size is 4kB}.

\subsection{Impact of Associativity on AMAT and MPKI}
\label{sec:amat}

We begin by assessing the associativity needs and benefits of L1
caches. To do this, we consider Average Memory Access Time (AMAT), the
most basic metric to analyze memory system performance. AMAT is
defined as:

\begin{equation}
AMAT = Hit Time + Miss Rate * Miss Penalty
\end{equation}

AMAT accounts for the effect of memory level optimizations across the
memory hierarchy. We first study the impact of cache associativity on
a key variable determining AMAT, the miss rate of the L1 cache.
Figure~\ref{fig:mpki} plots the MPKI averaged across a suite of big
data applications (see Section~\ref{sec:os_support} for workload
details) as a function of cache associativity for increasing cache
sizes.  Since we target L1 caches, we focus on relatively smaller 16kB
to 128kB capacities.  We observe that increasing associativity beyond
4 does not result in any noticeable reduction in miss rates.  This is
not just an artifact of our particular workloads, but a more
fundamental observation since the L1s are very small and service
requests only from one (or 2-4 if SMT) thread running over the core; a
low associativity is enough to reduce conflict misses, after which the
L1 is fundamentally limited by capacity misses~\cite{ovc}. This is
unlike LLCs which are orders of magnitude larger and are shared by
multiple cores and typically require large 8-16 way associativities to
mitigate conflicts in the set index bits among requests from the
various cores.
 
Unfortunately, modern L1 caches use associativities larger than 4 due
VIPT constraints.  While this does not reduce miss rates, it increases
hit time significantly, as we discuss next.

\subsection{Cache Access Time and Energy}
\label{sec:cacti_acctime_energy}

We performed a comprehensive study of how cache access latency and
lookup energy vary as a function of associativity across several cache
sizes using Cacti 6.5 \cite{cacti6.5}, at a 32nm node\footnote{This is
  the most advanced node that Cacti 6.5 supports. We also validated
  these trends using a 32nm ARM SRAM compiler, but cannot report the
  exact latency/energy numbers due to foundry confidentiality
  issues.}.  L1 caches are tightly coupled to the CPU pipeline, and
need to be optimized for both latency and energy.  Keeping this in
mind, we used the high-performance ("itrs-hp") transistor models, and
a parallel data and tag lookup for faster access.  For each design
point, Cacti to optimizes access time, cycle time, dynamic power, and
leakage power.

Figure~\ref{fig:acc_time} plots the absolute access latency as
associativity is increased from direct-mapped to 32-way for cache
sizes varying from 16kB to 128kB.  For each cache size, we observe
access latency to be somewhat flat (around 0.5ns for 16kB and 32kB,
and 1ns for 64kB and 128kB) till a low associativity of 4, beyond
which there is a steep increase of around 85\%
on average at 8-way, 
16-way, and 32-way.

A similar trend is also reflected in the total access energy graph in
Figure~\ref{fig:acc_energy}.  The total energy increases
monotonically, as more ways need to be read out and compared.  There
is a steep increase in both dynamic and leakage energy at 16-way and
32-way, as Cacti tries to use larger cells to optimize for meeting a
tight delay constraint. Dynamic energy shoots up by 45\%, while
leakage shoots up by 6$\times$ \footnote{At 32-way, 80\% of the energy
  is reported to be in leakage. This is due to a large number of
  standard cells required to implement the highly associative muxes,
  while meeting a tight timing constraint that we specified to
  Cacti. The leakage component could be reduced using advanced circuit
  techniques or relaxing the timing - which is orthogonal to this work
  - but the overall trend of higher lookup energy with more
  associativity would still remain.}.

These trends of increasing access time and energy are an intrinsic
artifact of cache associativity, and very similar graphs can also be
found in Hennessy and Patterson's 5th
edition~\cite{hennessy:architecture}.  These trends suggest that a low
4-way associative cache is best suited in terms of both access time
and energy across the given cache sizes.  This is a fundamental
observation.  The exact associativity at which the latency and/or
energy shoots up may vary based on the technology node, cache
micro-architecture, and on the tight timing or energy constraint that
is being optimized for. But we expect there to be a low associativity
number beyond which a multi-way lookup is going to fail to meet timing
(thus requiring an additional cycle for lookup) or the energy budget.
As benefits from technology scaling have plateaued, a fairly similar
behavior is expected at advanced nodes as well.

Thus L1 caches today are in a catch-22 situation - on one hand higher
associativity is required for supporting parallel TLB and L1 accesses
via VIPT (Section~\ref{sec:vipt}); on the other hand higher
associativity does not translate to increased hit-rates
(Section~\ref{sec:amat}), and in fact increases cache access time and
energy (Section~\ref{sec:cacti_acctime_energy}).  This makes L1 design
extremely challenging going forward, since on one hand future
workloads would have larger working sets~\cite{ferdman:cloudsuite,
  wang:bigdatabench, thomas:cortexsuite, ovc} that would benefit from
larger L1s, while on the other chips are highly power
constrained~\cite{esmailzadeh_dark_silicon}.  There are only two
solutions today; either use VIVT caches which require
complex synonym management and are not used in mainstream systems, or
use way prediction~\cite{sleiman2012embedded}, which is popular for
instruction caches~\cite{powell2001reducing} but harder to get right for data
caches.

We offer an alternate lightweight solution that provides the hit time
and energy benefits of lower associativity, without the challenges of
synonyms or a predictor\footnote{In fact, our solution can be
  augmented with way prediction to provide even more benefits, as we
  show later.}. We argue that the rigid assumption made by VIPT caches
may not be appropriate in modern systems, as we discuss next.

\begin{figure*}[tbh]
\vspace{-4mm}
\setlength{\abovecaptionskip}{0pt}
\setlength{\belowcaptionskip}{0pt}
\begin{center}
\includegraphics[width=0.95\linewidth, height=4.2cm]{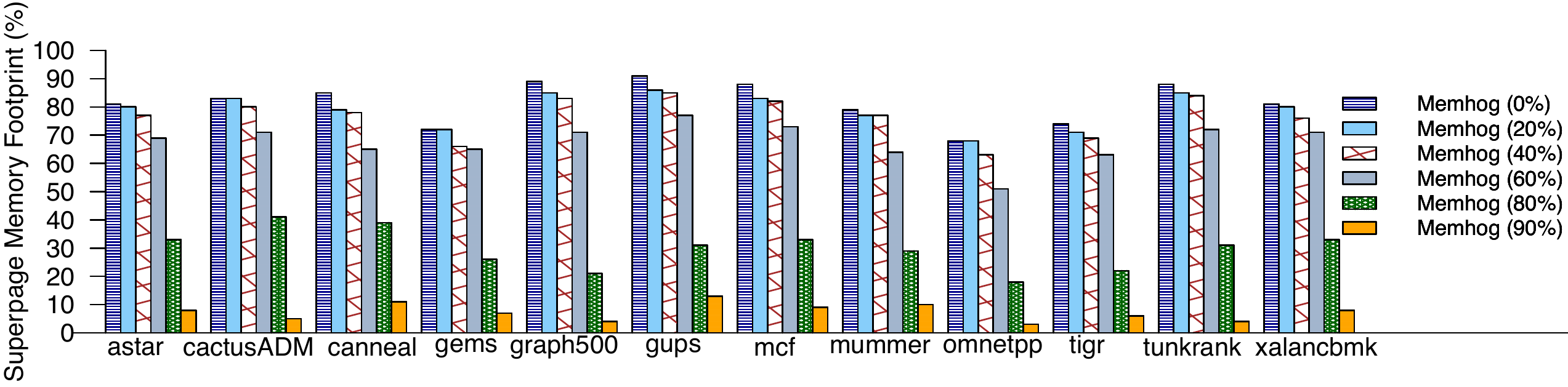}
\caption{\small \bf Fraction of total memory footprint allocated with 2MB
  superpages on a real 32-core Intel Sandybridge system with 32GB of
  memory. We show how superpage allocation varies as memory is
  fragmented with the {\sf memhog} workload.}
  \vspace{-2mm}
\label{fig:memhog}
\end{center}
\end{figure*}

\subsection{Superpages in Modern Systems}
\label{sec:os_support}

Our main observation is that real-world systems often use superpages
today.  The advantage with superpages is the availability of more bits
for virtual indexing, thus potentially increasing the number of sets
in a cache.

\vspace{2mm}\noindent{\bf Big data workloads:} A large class of
workloads today have large memory footprints, and stress virtual
memory heavily.  In this work, we use a suite of workloads from
CloudSuite~\cite{ferdman:cloudsuite}, Spec~\cite{spec} and
PARSEC~\cite{parsec} that have shown significant improvements in TLB
and system performance when using superpages across multiple
studies~\cite{pham:lightweight, ferdman:cloudsuite, gaud:large,
  gandhi:agile, gandhi:efficient, gorman:performance}. As such,
superpages are increasingly common \cite{pham:lightweight,
  gorman:performance, kwon:coordinated}.

\vspace{2mm}\noindent{\bf OS support for multiple page sizes:} Given
the increasing use of superpages, modern architectures support
multiple superpage sizes apart from the base page size. For example,
Linux, FreeBSD, and Windows all support not only 4KB base pages, but
also 2MB and 1GB superpages for x86 systems. Similarly, OSes for ARM
systems use 4KB base pages, but also 1MB and 16MB superpages. All page
sizes are useful. OSes allocate superpages to improve TLB hit rates,
and occasionally to reduce page faults (see past work for more details
\cite{navarro:practical, talluri:surpassing}). In contrast, OSes use
base pages when they desire finer-grained memory protection
\cite{witchel:modrian}. In fact, the benefits of multiple page sizes
have become so apparent that there is active work today on extending
superpage support on mobile OSes like Android \cite{dong:shared}. This
work targets server system so we omit a discussion of mobile
architectures; however, we expect that {\sf VESPA's} benefits may
likely extend to mobile architectures eventually as they adopt
superpages more aggressively.

\vspace{2mm}\noindent{\bf Superpages in long-running systems:} We
demonstrate that superpages are common in most server-class systems
today, even in the presence of modest to extreme memory fragmentation.
We ran all our applications on a real 32-core Intel Sandybridge
server-class system with 32GB of main memory, running Ubuntu Linux
with a v4.4 kernel.  The system had been active and heavily loaded for
over a year, with user-level applications and low-level kernel
activity (e.g., networking stacks, etc.). To further load the system,
we ran a memory-intensive workload called {\sf memhog} workload in the
background.  {\sf memhog} is a microbenchmark for fragmenting memory
that performs random memory allocations, and has been used in many
prior works on virtual memory \cite{CoLT, pham:lightweight,
  pham:clustering}. For example, {\sf memhog (50\%)} represents a
scenario where {\sf memhog} fragments as much as half of system
memory.  We enabled Linux's transparent hugepage support
\cite{THP_redhat} that attempts to allocate as much anonymous heap
memory with 2MB pages as possible. Naturally, the more fragmented and
loaded the system, the harder it is for the OS to allocate superpages.

Figure \ref{fig:memhog} plots the fraction of the workload's memory
footprint allocated to superpages.  Fundamentally, it reveals that
modern systems frequently use superpages today. When fragmentation is
low (i.e., {\sf memhog (0-20\%))}, 65\%+ of the memory footprint is
covered by 2MB superpages, for every single workload. In many cases,
this number is 80\%+. Furthermore, even in systems with reasonable
amounts of memory fragmentation (i.e., {\sf memhog (40-60\%))},
superpages continue to remain ample. This is not surprising, since
Linux -- and indeed, other OSes like FreeBSD and Windows -- use
sophisticated memory defragmentation logic to enable superpages even
in the presence of non-trivial resource contention from co-running
applications. It is only when contention increases dramatically ({\sf
  memhog(80-90\%)} that OSes struggle to allocate
superpages. Nevertheless, even in the extreme cases, some superpages
are allocated.

We also studied FreeBSD and Windows and we found that 20-60\% of the
memory footprint of our workloads, running on long-running server
systems which have seen lots of memory activity, are covered by
superpages.  On average, the number is roughly 48\%. Overall, this
data suggests that OSes not only support, but actually {\it use}
multiple page sizes with even modest and extreme memory fragmentation.

\vspace{2mm}\noindent{\bf Hardware support for multiple page sizes -
  Split TLBs:} Though many hardware components interact with the
virtual memory system, the two most important ones are the TLBs and
the L1 caches. Modern processors use a combination of TLBs to cache
virtual-to-physical translations.

Multiple page sizes impose an important challenge on TLB design. As
TLBs often consume 13-15\% processor power \cite{ovc}, vendors usually
use set-associative TLBs. It is, however, challenging to design a
single set-associative TLB for all page sizes.  Recall that TLBs are
looked up using the virtual page number (VPN) of a virtual
address. Identifying the VPN, and hence the set-index bits, requires
masking off the page offset bits.  This is a chicken-and-egg problem -
the page-offset requires knowledge of the page- size, which is
available only {\it after} TLB lookup \cite{cox:efficient}.

To get around this problem, vendors use {\it split TLBs} at the L1
level (there are higher-latency workarounds for L2) for different page
sizes \cite{papadopoulou:prediction, seznec:skew}.  For example, Intel
Sandybridge systems use 64-entry L1 TLBs for 4KB pages, 32-entry L1
TLBs for 2MB pages, and 8-entry L1 TLBs for 1GB
pages~\cite{sandybridge}.  Memory references probe all TLBs in
parallel. A hit in one of the TLBs implicitly identifies the page
size. Misses probe an L2 TLB, which can occasionally support multiple
page sizes. Irrespective of the microarchitectural details, TLB
hierarchies today are designed to operate harmoniously with multiple
page sizes.

This is in stark contrast with VIPT L1 caches, as we have detailed
thus far. VIPT L1 caches do support multiple page sizes, but not as
efficiently as possible. In other words, since VIPT is conservatively
tuned to support the page offset width of the base page size, it
needlessly penalizes access to superpages. {\sf VESPA} attacks exactly
this problem.

\section{VESPA Microarchitecture}\label{sec:VESPA}

\begin{figure*}[tbh]
\setlength{\abovecaptionskip}{0pt}
\setlength{\belowcaptionskip}{0pt}
\includegraphics[width=1\linewidth]{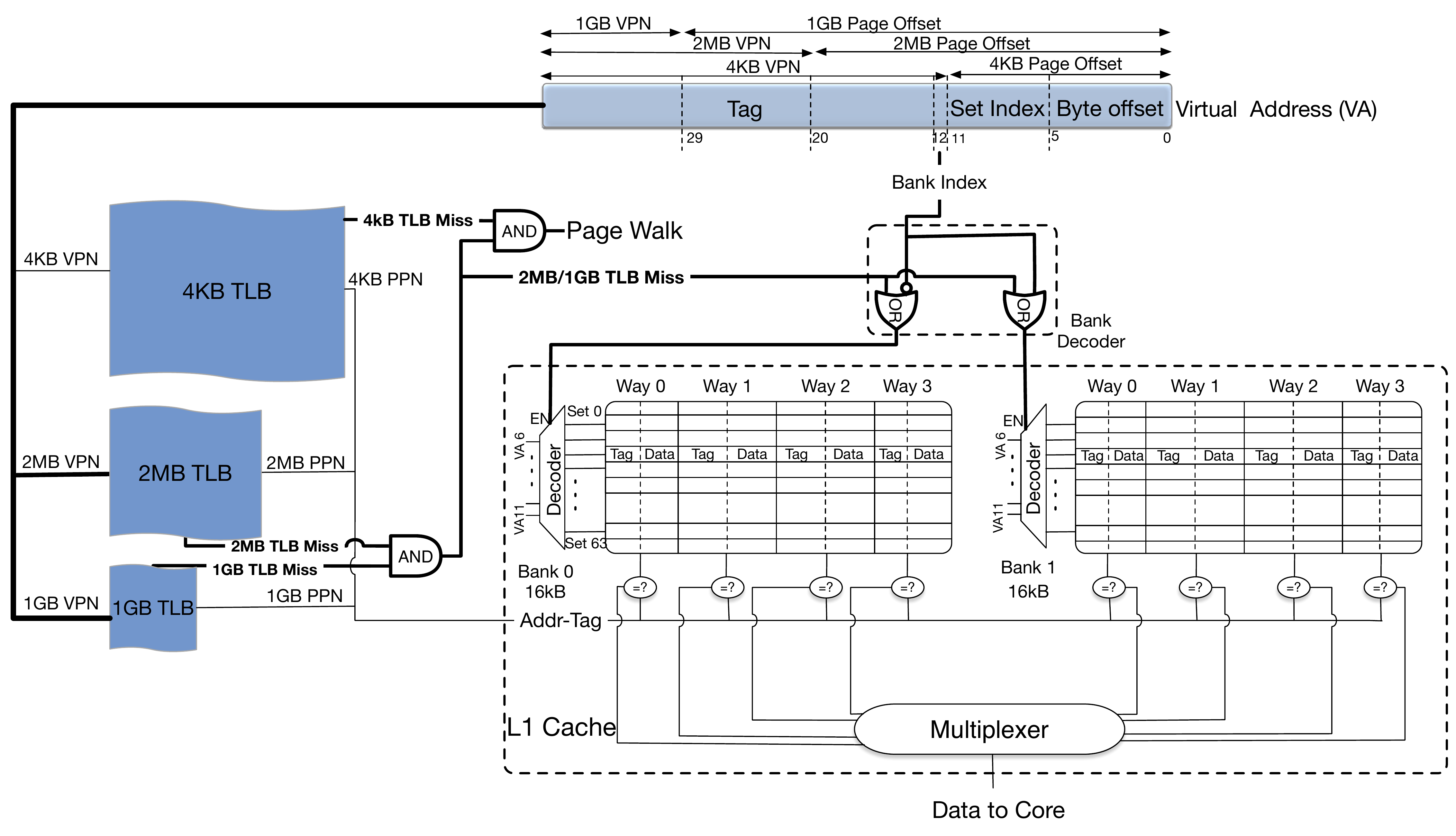}
\caption{\bf \small VESPA Microarchitecture for a 32kB L1 Cache.}
\label{fig:micro_arch}
\end{figure*}

We have shown that L1 caches often achieve their best
energy-performance points when they use relatively low
set-associativity. Unfortunately, traditional VIPT constraints often
preclude this possibility, as they are unable to realize high set
counts (e.g., 64+ sets in x86 systems) and instead require high
associativity.  In this work, we modify the L1 cache hardware to
support a new flavor of VIPT without these constraints by embracing
the opportunity presented by superpages.  We fulfill three design
steps: \textcircled{\small a} we judiciously modify existing L1 cache
hardware and align it with the existing TLB hierarchy,
\textcircled{\small b} we explore optimizations for better cache
insertion policies and scalability; and \textcircled{\small c} we
consider the implications of these changes on system-level issues like
cache coherence, and OS-level paging.

{\it We showcase these steps for a 32KB, 8-way L1 cache operating at
  1.33 GHz for x86 with 4KB base pages, and 2MB/1GB superpages.}
However, our approach is equally applicable to other L1 cache sizes,
as highlighted by our evaluations, and other architectures with
different page sizes.

\subsection{Hardware Augmentations}
\label{sec:hardware-augmentations}

{\sf VESPA} operates as follows. When a memory reference is to a base
page, {\sf VESPA} looks up the same number of ways as a traditional
VIPT cache. However, when the reference is to a superpage, {\sf VESPA}
needs to look up fewer cache ways, saving on both access latency and
energy.

  \subsubsection{L1 cache microarchitecture}
  \label{sec:microarch}
{\sf VESPA} uses a banked-microarchitecture for the L1 cache, i.e.,
each set is divided into multiple banks.  Each bank is organized with
a target associativity that is chosen for its desirable latency and
energy spec.  Recall that the page-offset bits remain the same in both
the VA and the PA.  {\sf VESPA} exploits the fact that superpages
increase the number of page-offset bits within the virtual address
(VA) (for instance 21 bits for 2MB pages and 30 bits for 1GB pages, as
Figure~\ref{fig:micro_arch} shows).  Thus, for superpages, it uses
some of the additional page-offset bits as a {\it bank\_index} to
index into one of the banks of the set (which is selected by the
set\_index, as is usual) and only needs to lookup the {\it ways within
  the bank}.

Accesses to base pages require a lookup of all banks (i.e., {\it all
  the ways in the set}) like a traditional VIPT design, since the {\it
  bank\_index} bits are now part of the VPN which would change in the
PA after address translation.

The banked implementation can be used for the baseline VIPT caches as
well, like AMD's Jaguar
does, and is an implementation
choice.  But it does not directly provide any latency benefit over a
non-banked one since the ways in all banks need to be looked up
anyway; for instance for the 32kB design, the lookup of all ways takes
2 cycles - either 4-ways serially each cycle, or 8-ways in parallel
across both cycles.  The total access energy for a serial vs parallel
lookup was also found to be similar for a 8-way cache; but might make
a difference at very large associativities.

Figure~\ref{fig:micro_arch} shows the microarchitecture of a 32kB cache, which requires 8-ways for a VIPT access.
In {\sf VESPA}, we partition the 8-way set into two banks of 4-ways each; 
based on the latency and energy characterization presented earlier in Figures~\ref{fig:acc_time} and~\ref{fig:acc_energy}.
We introduce a {\it bank decoder} to index into one of the banks, using the {\it bank\_index} (bit 12 of the virtual address).
We present the policy for indexing into one of the banks (for superpages) versus reading all banks (for base pages) next.
Table~\ref{table:config} presents the access latencies in the cache for baseline and superpages across various cache sizes 
and clock frequencies, to point to the robustness of this idea.

\begin{table*}[pht]
\small
\caption{\small \bf Anatomy of a Lookup} \label{table:anatomy}

\centering
\begin{tabular}{| l | l | l | l | l | l |}
\hline
{\bf Page Size} & {\bf TLB} & {\bf Cache} & {\bf Cycle 1} & {\bf Cycle 2} & {\bf Savings}\\
&  &  &  &  &{\bf  over}\\	
&  &  &  &  &{\bf Baseline}\\	
\hline
2MB/1GB & Hit & Hit & Bank lookup using bank\_index (bit 12 of the VA).& Not Required. & Latency\\
&  &  &  Tag matches. This is the same case as a traditional & & +\\
&  &  &   VIPT for a 4-way  cache. & & Energy\\
\hline
2MB/1GB & Hit & Miss & Bank lookup using bank\_index (bit 12 of the VA).& &\\
&  &  & Tag mismatch triggers cache miss. & Not Required. & Energy\\
\hline
2MB/1GB & Miss & * & Bank lookup using bank\_index (bit 12 of the VA).& Other bank is read. 4kB TLB miss&\\
&  &  &   2MB/1GB TLB miss signal triggers a read of the & triggers Level-2 TLB (if present) &\\
&  &  &   remaining 4-ways of the adjacent bank & lookup which may trigger a page & None\\ 
&  &  &   (i.e., assume a 4kB page).& table walk. &\\ 
\hline
4kB & Hit & Hit & Bank lookup using bank\_index (bit 12 of the VA). & Other bank is read. Tag matches. & \\
&  &  &   The 2MB/1GB TLB miss signal triggers a read  &This is the same case as a traditional & None\\ 
&  &  &   of the remaining 4-ways of the adjacent bank.& VIPT for a 8-way set associative cache & \\ 
\hline
4kB & Hit & Miss & Appropriate bank is looked up using the & Other bank is read. Tag mismatch & \\
&  &  &    bank\_index (bit 12 of the VA).& triggers cache miss. & None\\
&  &  &   The 2MB TLB miss signal triggers a read of the & &\\
&  &  &    remaining 4-ways of the adjacent bank. & &\\
\hline
4kB & Miss & * & Appropriate bank is looked up using the & Other bank is read. 4kB TLB miss &\\
&  &  &   bank\_index (bit 12 of the VA).& triggers Level-2 TLB (if present)  & None\\
&  &  &   The 2MB TLB miss signal triggers a read of the & lookup which may trigger a page &\\
&  &  &   remaining 4-ways of the adjacent bank. & table walk &\\
  \hline
\end{tabular}
\label{table:lookup_anatomy}

\end{table*}

 \subsubsection{TLB-L1 cache interface}
 \label{sec:TLB-L1-interface}
{\sf VESPA} needs to infer, from the virtual memory lookup address,
whether a reference is to a superpage (2MB/1GB) or a base page (4kB),
and accordingly lookup the right bank or all banks respectively.  One
option is to look up the TLB for the virtual-to-physical translation
to determine the page size and then look up the L1 cache. Naturally,
this option is a non-starter as serially looking up the TLB and L1
cache excessively degrades performance and is what VIPT caches want to
avoid in the first place.

Instead, recall that CPUs use dedicated {\it split TLBs} for different
page sizes \cite{papadopoulou:prediction, Intel_skylake, IBM_P8,
  cox:efficient}, as discussed in Section~\ref{sec:os_support}. We
piggyback on this design.  The separate TLBs are sized as per the page
size they support. In other words, TLBs for 2MB pages have {\it fewer
  entries} than TLBs for 4KB pages since each 2MB TLB entry covers a
larger chunk of the address space than each 4KB TLB entry. For
example, Intel Sandybridge processors use 2MB-page L1 TLBs with half
the number of entries as the 4KB-page L1 TLB. Further, each 2MB-page
TLB entry requires fewer bits to store the virtual and physical page
numbers than 4KB-page TLB entries. We find that 2MB-page L1 TLBs are
40\% smaller than 4KB-page L1 TLBs. 1GB-page L1 TLBs are even smaller.
Consequently, {\it superpage TLBs have much shorter access latencies
  than base page TLBs.}  Table~\ref{table:config} lists TLB access
latencies for varying sizes and clock frequencies.

The split TLBs with differing access times drive {\sf VESPA} as
follows.  All memory references look up the TLB hierarchy and L1 cache
in parallel. However, unlike conventional VIPT, {\sf VESPA} performs
the L1 cache lookup {\it speculating a superpage access}.  Therefore,
not only is a specific set chosen, so too is a specific bank within
that set, using the bank\_index bits.  In parallel, the 2MB/1GB-page
TLB lookup, which is faster than the 4kB lookup (e.g., half the time
at 1.33GHz), indicates whether the access is to a superpage.  If not,
i.e., the speculation failed, the L1 cache logic begins a lookup of
the remaining banks in the set. Accesses to lines on superpages thus
finish faster, while those for base pages takes the same time as
before.

\subsubsection{Anatomy of a Lookup}\label{sec:anatomy}

Table~\ref{table:lookup_anatomy} lists the cache lookup timeline 
on a case-by-case basis for a 32kB L1 at 1.33GHz on a x86 machine 
with 4kB base pages and 2MB/1GB superpages.
A superpage access takes 1-cycle in this design, and a base page takes 2-cycles.
The behavior for other configurations (Table~\ref{table:config}) 
and page sizes can accordingly be derived.

\subsubsection{Implementation Overheads}
\label{sec:implementation}
{\sf VESPA} uses set banking to support the ability to dynamically
change associativity from 8- to 4-way. Two modest hardware
enhancements are needed. The first is a bank decoder (two 2:1 OR gates
in Figure~\ref{fig:micro_arch}) placed before the banks. The second is
an extra 2:1 mux at the end to choose between the two banks.  Within
each bank, {\sf VESPA} needs a 4:1 mux instead of the
8:1 mux used by baseline VIPT for the entire set.  We updated the
Cacti cache model to implement these changes, using the 32nm
standard-cells used internally for implementing the decoder and the
muxes.  We observe less than 1\% increase in access time, which does
not our affect cycle time at 1.33GHz as it is within the timing margings in the
design.  The lookup energy for a 4-way access in {\sf VESPA}
increases by just 0.41\%, which is still 39.43\% lower than that for
 8-way
access in the baseline.

\subsection{Design Optimizations}
\label{sec:optimizations}

\subsubsection{Cache Line Insertion Policy}
\label{sec:VESPA line insertion}
Since {\sf VESPA} dynamically switches the L1 cache from 8- to 4-way,
there can be two variants of traditional cache line insertion
policies.

\vspace{1mm}\noindent{\textcircled{\small a} \bf 4way-8way insertion
  policy:} On a cache line miss from a superpage (see
Table~\ref{table:lookup_anatomy}), the replacement victim line is
chosen from the same bank using an LRU policy.  However, if there is a
cache line miss from a base 4KB page, the replacement victim is chosen
across both banks, by following an LRU policy.  Thus, {\sf VESPA}
behaves like a 4-way associative cache for superpages and a 8-way
associative cache for base pages from an insertion policy perspective.

\vspace{1mm}\noindent{\textcircled{\small b} \bf 4way insertion
  policy:} On a cache line miss from either a superpage or a base
page, the line is installed in the bank specified by the bank\_index
bits from the physical address (which is available post TLB lookup).
The replacement victim is chosen using an LRU policy from the same
bank.  The 4way policy uses a local replacement policy within the 4
ways of the concerned bank, instead of a global replacement within 8
ways of the original set, irrespective of page size.

We decided to use the 4way insertion policy in {\sf VESPA} for 4
reasons.  \textcircled{1} Correctness: There may be cases where a page
is mapped both as a base page and a superpage. A 4way-8way policy
might lead to the same line getting installed twice in the cache; a
uniform policy for both base and superpages avoids this problem.
\textcircled{2} Energy: The LRU policy is simpler and saves energy on
each cache-line installation due to tracking and lookup of fewer ways.
\textcircled{3} Performance: As an academic exercise, we ran all our
experiments with both policies, and noticed only a 1\% difference drop
in hit rate with the 4way policy, in line with the earlier
observations in Figure~\ref{fig:mpki}.  \textcircled{4} Coherence
lookups: The 4-way policy reduces lookup time and energy for coherence
lookups, as we discuss later in Section~\ref{sec:system_issues}.

\subsubsection{Supporting Larger L1 Caches}\label{sec:extend_VESPA}

Current trends suggest that L1 cache sizes will grow beyond 32KB with
coming processor generations.  AMD's Jag\-uar uses 64kB caches, and
Excavator chips ~\cite{excavator} are expected to use 128KB L1
caches\footnote{There are no publicly released documents yet but tech
  websites suggest an implementation using 4 banks of 32kB caches}.
Increasing L1 size exacerbates the access latency and energy of VIPT
L1 caches, since they are usually grown by increasing associativity
(see Section~\ref{sec:vipt}).  Our detailed studies with Cacti 6.5
(Figures~\ref{fig:acc_time} and ~\ref{fig:acc_energy}) suggests even
worse latency and energy with higher associativity at 64KB and 128KB
cache sizes.  This makes {\sf VESPA} even more relevant and necessary
going forward.

\begin{figure}[!tb]
\includegraphics[width=0.9\linewidth]{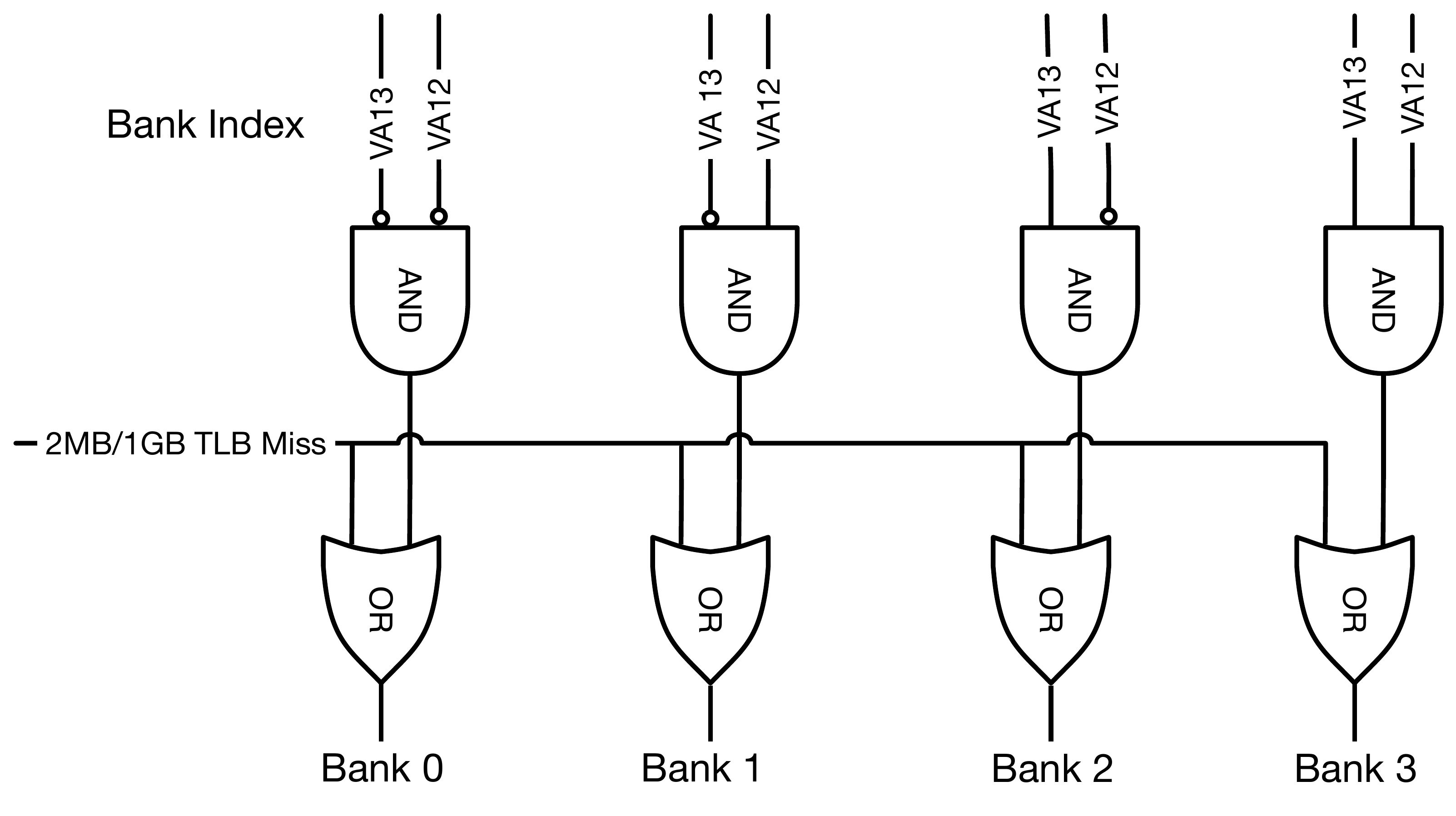}
\caption{{\bf \small Bank Decoder for 64KB VESPA.}}
\label{figure:64KB Micro-arch}
\end{figure}%

{\sf VESPA's} operation for larger L1 caches mirrors our description
of 32KB L1 caches. The difference is that the number of banks, each of
which is 16KB, increases.  This changes the bank decoder circuitry
(Figure~\ref{figure:64KB Micro-arch}).  We show the possible circuit
of a bank decoder for a 64KB {\em VESPA} in Figure \ref{figure:64KB
  Micro-arch}. Similarly, a bank decoder for a 128KB cache can be
built.  We choose 16KB as the bank size because our Cacti-based
analysis suggest that 4-way associativity remains optimal for even
64KB and 128KB L1 caches.  If this number happens to be different at
another technology node for a certain cache model, different bank
sizes can be accommodated, with the cache being built using multiples
of these banks.

One caveat in VIPT cache design is scaling of TLBs. As the cache size
grows, its access latency increases, and accordingly TLB sizes can
also be increased in terms of number of entries. This way, VIPT caches
can still perform parallel TLB and cache lookups.  In our 64KB and
128KB caches, the TLBs are sized accordingly to fulfill the above
mentioned condition. Our evaluations (Section~\ref{sec:eval}) show
even higher benefits in terms of energy and AMAT with {\sf VESPA}
as cache size increases.

\subsubsection{Way Prediction}
As VIPT caches are highly set-associative (Section~\ref{sec:vipt}),
{\sf VESPA} can essentially be viewed as a predictor for which subset
of ways (which we call a bank) to lookup.  {\sf VESPA}'s approach is
to predict all accesses to be superpage accesses, and access the
appropriate bank using the {\it bank\_index} bit(s), and access the
other banks on a mis-prediction (signaled by monitoring the TLB
hit/miss signals, as Table~\ref{table:lookup_anatomy} discussed).

An orthogonal approach that that is potentially symbiotic with {\sf
  VESPA} is the idea of {\it way-prediction} \cite{powell:reducing,
  faissal:embedded}. Way-prediction uses simple hardware to predict
which way in a cache set is likely to be accessed in the future. When
the prediction is accurate, access energy is saved without
compromising access latency, as the cache effectively behaves in a
direct-mapped manner. The key is accurate prediction, with past work
proposing several schemes like using MRU, the PC, or XOR functions to
achieve this \cite{faissal:embedded}. Naturally, predictor accuracy
can vary, with good results for workloads with good locality and poor
results for emerging workloads with poor access locality (e.g., like
graph processing). 

{\sf VESPA} presents an effective additional design point to
way-prediction. Crucially, {\sf VESPA} can improve not only access
energy but also access latency (which way-prediction does not target).
This is particularly important when way-predictor accuracy is
sub-optimal. Naturally, both approaches can be used in tandem; {\sf
  VESPA} can reduce the penalty of way-predictor mistakes by reducing
the number of cache ways that need to be looked up on a
misprediction. Conceptually, combining both approaches provides an
effective means of tackling L1 caches -- {\sf VESPA} always reduces
energy and latency for superpage accesses and way-prediction
effectively reduces the energy of base-page accesses for which the
predictor accurately guesses the required way. We evaluate the
confluence of these designs in subsequent sections.

\subsection{System-Level Issues}
\label{sec:system_issues}

{\sf VESPA} has several interesting implications on system-level
issues involving multi-core interactions and the software stack. We
discuss some of these interactions here.

\vspace{2mm}\noindent{\bf Cache coherence:} One may initially expect
cache coherence to be unaided by {\sf VESPA}. However, the choice of
insertion policy (see Sec. \ref{sec:VESPA line insertion}) affects
coherence lookup (invalidations/probes) access time and energy from
the L2/directory. Consider the 4way-8way insertion policy. Coherence
lookups for cache lines residing in superpages only need to search in
the 4 ways of the appropriate bank, as the coherence request comes
with the physical address.  However, for cache lines on a 4KB page,
coherence lookups at L1 need to search in all the 8 ways by activating
both banks. Here VESPA saves energy for all superpage coherence
requests and does no worse for baseline 4KB page coherence requests.

However, if we implement the 4way insertion policy, the correct bank
can be accessed using the bank\_index bits of the physical address for
all requests, whether on base pages or superpages.  This saves energy
for {\it every} coherence lookups, irrespective of page size.  As
mentioned in Section~\ref{sec:VESPA line insertion}, there is minimal
change in the hit rate for L1 caches with this policy and it does not
change the AMAT of the system.  Hence we use this policy in rest of
our paper.

\vspace{2mm}\noindent{\bf Page table modifications:} Important
system-level optimizations like copy-on-write mechanisms, memory
deduplication, page migration between NUMA memories, checkpointing,
and memory defragmentation (to generate superpages) rely on modifying
page table contents. Some of these modifications can cause superpages
to be converted into base pages (or vice-versa). In these cases, while
the physical addresses of data in these pages remains unchanged, they
must now be correctly treated by {\sf VESPA} as residing in base pages
instead of superpages. There are two cases to consider.

First, suppose that a superpage is broken into constituent base
pages. We must ensure that L1 cache lines that belonged to the
superpage are correctly accessed. Fortunately, this is simple. {\sf
  VESPA} looks up more L1 cache banks for data mapping to base pages
than superpages; in fact, accesses to base pages automatically also
look up the bank that the superpage would have previously been allowed
to fill. Therefore, there are no correctness issues when transitioning
from 2MB/1GB pages to 4KB pages.

Second, several base pages may together be promoted to create a
superpage.  Since {\sf VESPA} probes fewer banks, it is possible a
line from one of the prior base pages may be cached in a bank that is
no longer probed. Naturally, this is problematic if that line
maintains dirty data. While several solutions are possible, we use a
simple - albeit heavyweight - one. When the OS promotes base pages to
a superpage, it has to invalidate all the base page translation
entries in the page table. For correctness reasons, OSes then executes
an instruction (e.g., {\sf invlpg} in x86) to invalidate cached
translations from the TLBs. These instructions are usually
high-latency (e.g., we have designed micro benchmarks that estimate
this latency to be 150-200 clock cycles, consistent with measurements
made by Linux kernel developers). We propose overlapping extending
this instruction so that it triggers a sweep of the L1 cache, evicting
all lines mapping to each invalidated base page. In practice, we have
found 150-200 cycles more than enough to perform a full cache
sweep. Finally, we model such activities in our evaluation
infrastructure and find that page table modifications events only
minimally affect performance.

\vspace{2mm}\noindent{\bf TLB-bypassing:} Occasionally, there may be
accesses to physical data without a prior virtual-to-physical
translation. For example, page table walkers are aware of the physical
memory address of the page tables and do not look up the TLB. The OS
also accesses several data structures in a similar way. Since these
data structures do not reside in virtual pages, {\sf VESPA} can
handle their L1 cache lookups similar to base page or superpages. To
promote good performance and energy efficiency, their lookups mirror
the superpage case.

\begin{center}
\small
\begin{tabular} {|m{9em}|m{4.4cm}|}
\hline
{\bf CPU}& In-order, x86, 1.33GHz\\
\hline
{\bf L1 Cache} & Private, Split Data \& Instruction\\ 
\hline
{\bf L2 cache (Unicore)} & Unified, 1MB, 16-way\\
\hline
{\bf L2 cache (Multicore)} & Unified, 8MB, 16-way\\
\hline
{\bf DRAM}& 4GB, 51ns access latency\\
\hline
{\bf Technology}& 32nm\\
\hline
\end{tabular}
\captionof{table}{\bf \small System Parameters}
\label{table:system}
\end{center}

\section{Evaluation}
\label{sec:eval}

\subsection{Target System}


We evaluate {\sf VESPA} within a target x86 system described in
Table~\ref{table:system}, and compare its performance and energy
against a baseline VIPT ({\bf BaseVIPT}) cache.  To demonstrate the
robustness and benefits of our scheme with different processor
frequencies and L1 cache sizes, we evaluated {\sf VESPA} under various
configurations.  We target 32kB, 64kB, and 128kB caches, with
frequencies of 1.33GHz, 2.80GHz (e.g., AMD Phenom) and 4.00GHz (e.g.,
Intel Skylake).  Table~\ref{table:config} shows the access latency of
the L1 caches under each configuration: } The L2 cache is the Last
  Level Cache (LLC) in our system.

\vspace{2mm}
{\bf \noindent Instruction Caches.}  OSes like Linux do not currently
have superpage support for instruction footprints.  This has
historically been because instruction footprints have generally been
considered small, and hence a poor fit for 2MB/1GB superpages.  Thus
our studies focus on L1 data caches.  However, we believe that this
could change with the advent of server-side and cloud workloads
\cite{ferdman:cloudsuite, wang:bigdatabench} that use considerably
larger instruction-side footprints.  In this context, {\sf VESPA}
would support and benefit L1 instruction caches too.

\vspace{2mm}
{\bf \noindent Performance and Energy Metrics.}
To measure the access time and total energy of the L1 cache, we 
use Cacti 6.5\cite{cacti6.5} with the configuration mentioned in
Section \ref{sec:cacti_acctime_energy}. Since we optimize L1 lookups  
in terms of both latency and energy, we report AMAT (Average Memory 
Access Time) as our metric of performance and total energy of the L1 cache 
as our metric of energy efficiency. 

\subsection{Single-Core Performance and Energy}

\subsubsection{Methodology and Workloads}

\begin{figure}[tb]
\setlength{\abovecaptionskip}{0pt}
\setlength{\belowcaptionskip}{0pt}
\begin{center}
\includegraphics[width=0.95\linewidth, height=4cm]{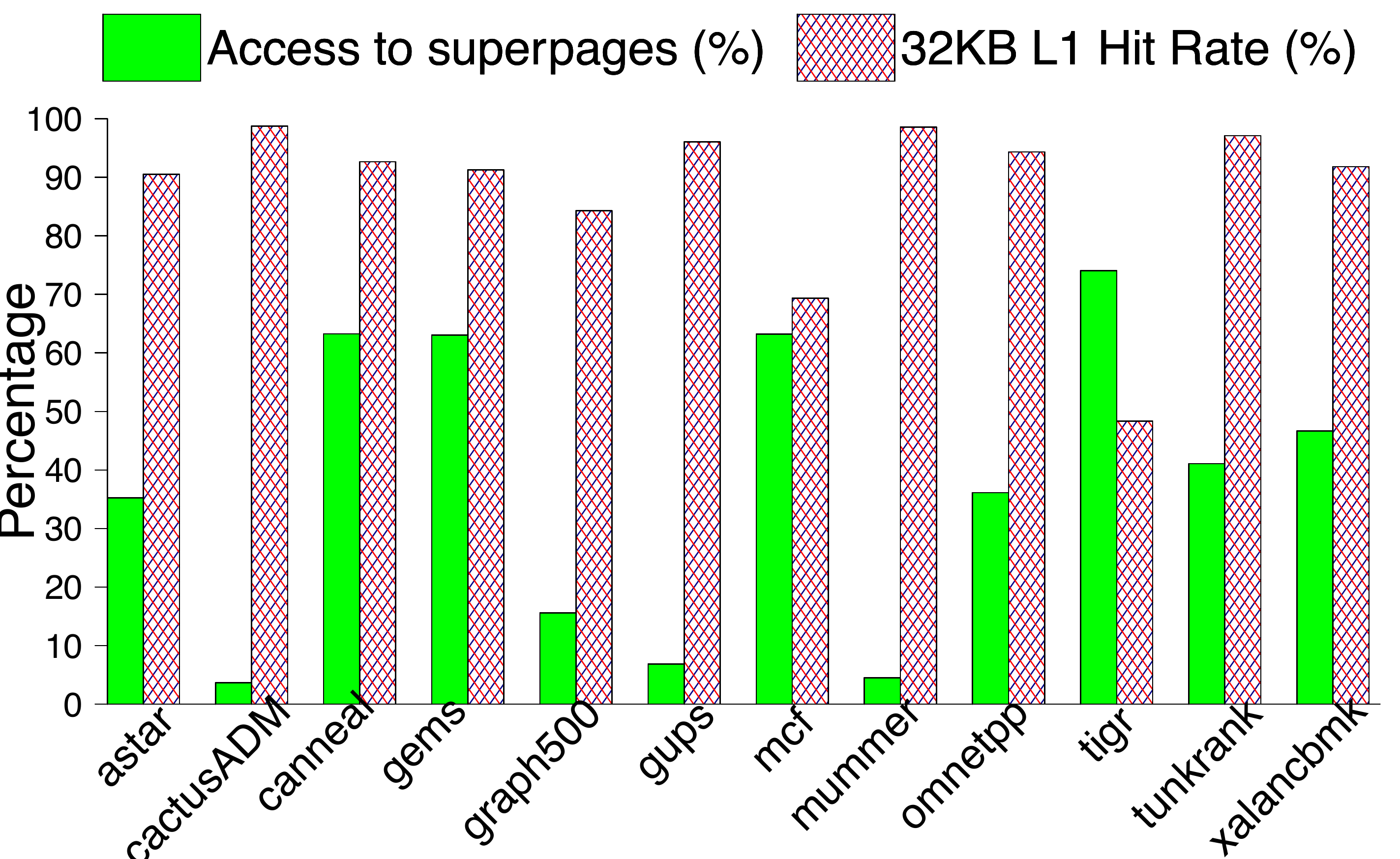}
\vspace{-2mm}
\end{center}
\caption{\bf \small Percentage of accesses to superpages, and L1 hit rates.} 
\label{fig:superpage-access_L1-hit-rate}
\end{figure}

\begin{figure}[tb]
\setlength{\abovecaptionskip}{0pt}
\setlength{\belowcaptionskip}{0pt}
\includegraphics[width=\linewidth, height=4cm]{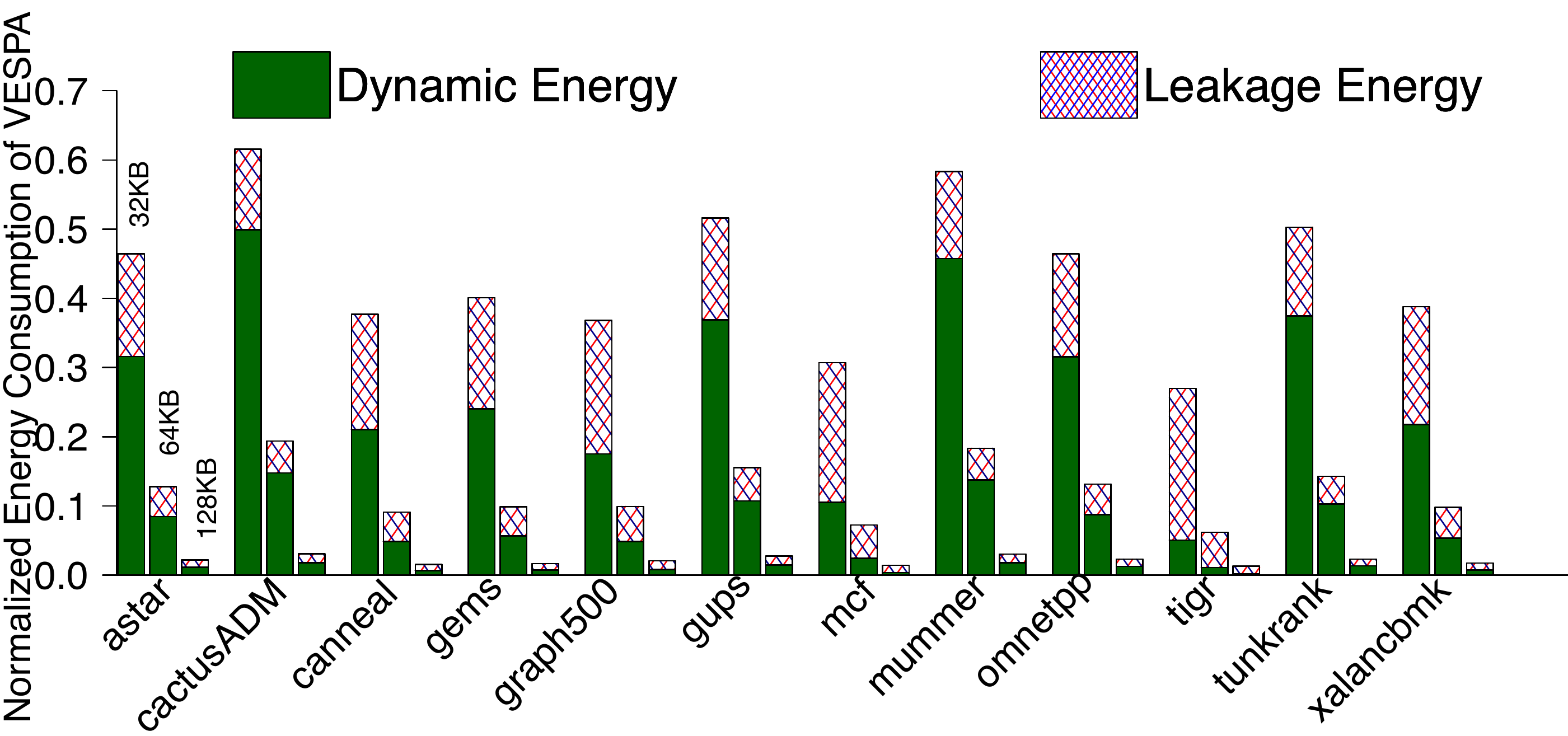}
\caption{\bf \small Normalized total L1 energy improvement provided by {\sf VESPA} over the baseVIPT cache
 for 32kB, 64kB and 128kB L1 cache sizes respectively.}
\label{fig:leakage_DynamicEnrgyWrkld}
\end{figure}

\begin{figure*}[tbh]
\setlength{\abovecaptionskip}{0pt}
\setlength{\belowcaptionskip}{0pt}
\centering
\subfloat[\small Normalized AMAT improvement @ 1.33GHz]
{
\includegraphics[width=0.34\linewidth]{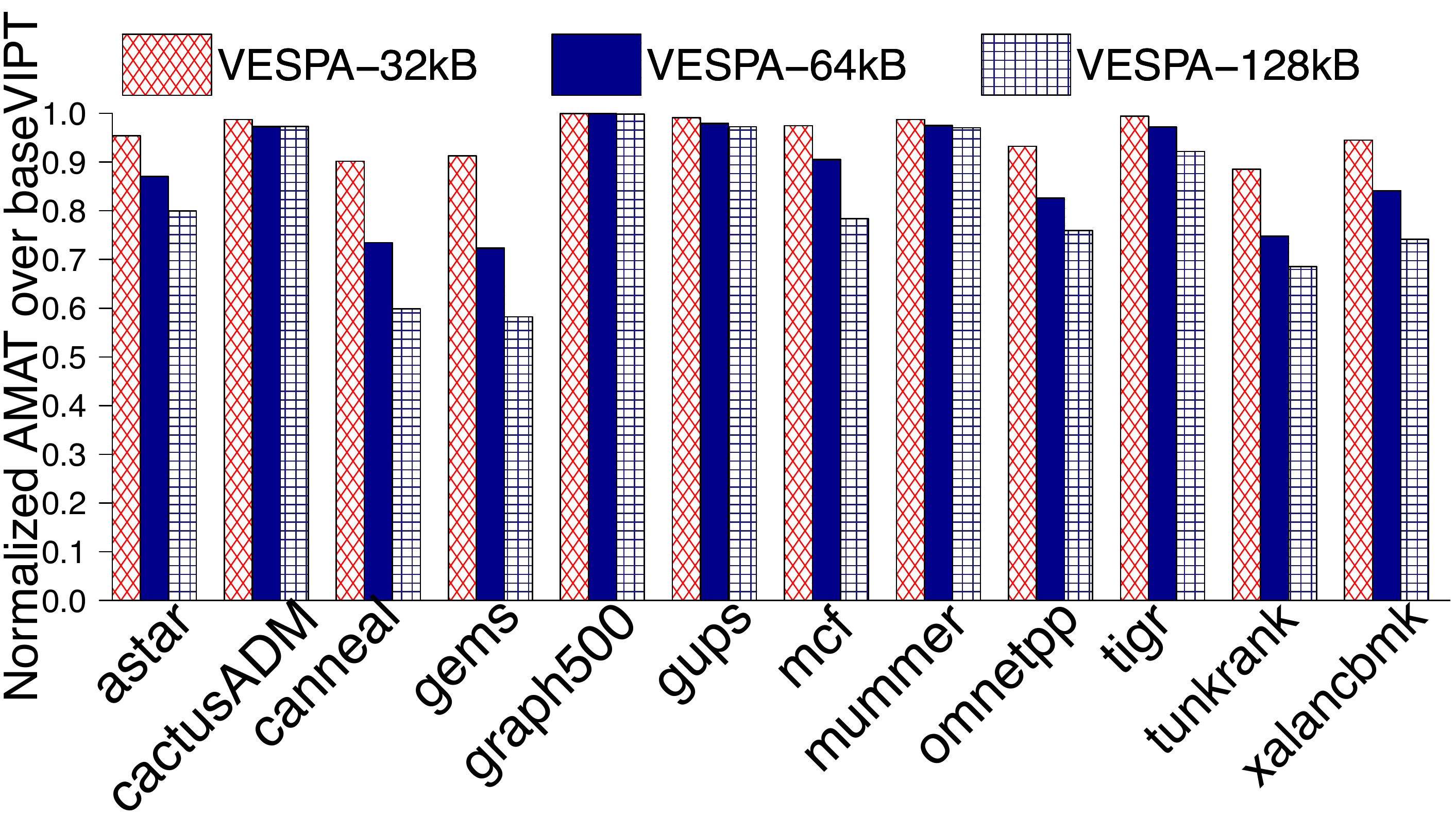}
\label{fig:amat_wrkld}
}
\subfloat[\small  Normalized AMAT improvement @ 2.80GHz]
{
\includegraphics[width=0.34\linewidth]{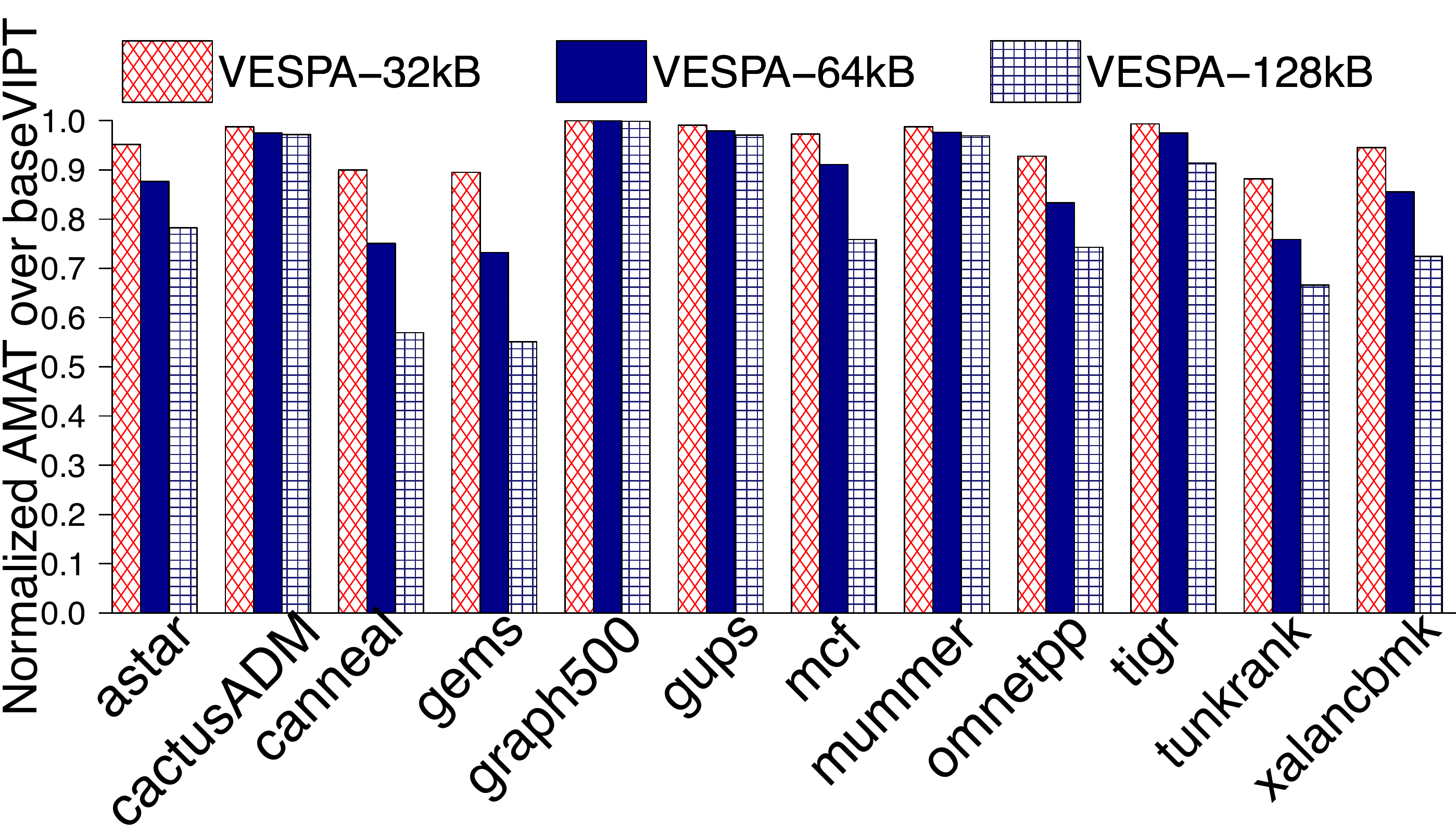}
\label{fig:amat_2.80GHz}
}
\subfloat[\small Normalized AMAT improvement @ 4.00GHz]
{
\includegraphics[width=0.34\linewidth]{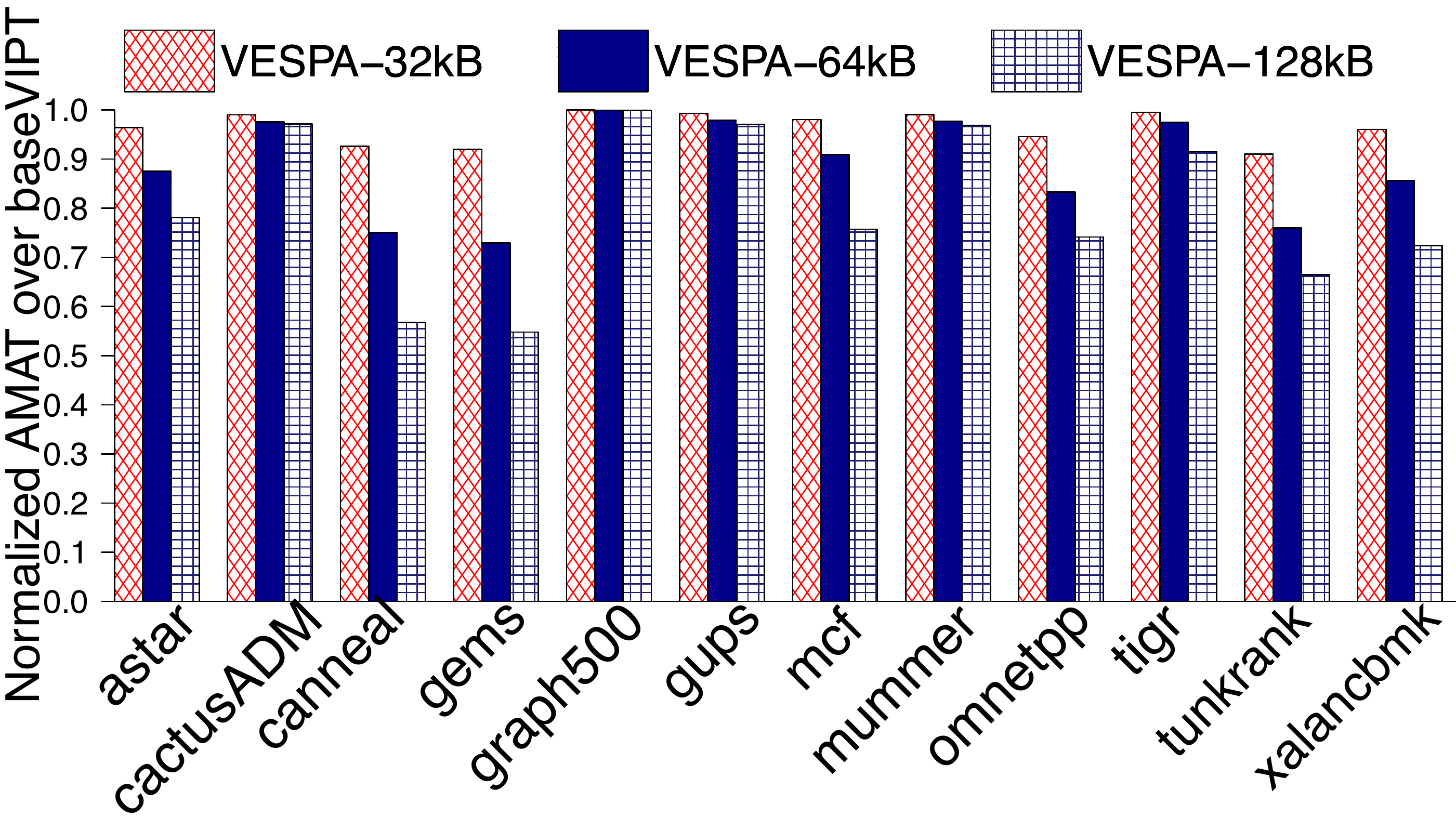}
\label{fig:amat_4.00GHz}
}
\caption{\small \bf Normalized AMAT improvement by VESPA at different frequencies for 32kB, 64kB and 128kB L-1 cache sizes respectively.
Each bar is normalized to its corresponding baseline VIPT cache for a given cache size.}
\label{fig:fragmentation_sweep}
\end{figure*}

Our single-core studies take a two-step approach, using detailed
memory tracing from a real 32-core Intel Sandybridge system with 32GB
of RAM and Ubuntu Linux (v4.4 kernel), as well as careful
simulation. We pick long-running systems with on-times of several
months to ensure that our system has the memory fragmentation and load
representative of server-class scenarios. We pick several workloads from
Spec~\cite{spec}, Parsec~\cite{parsec},
Biobench~\cite{jaleel:biobench}, and
Cloudsuite~\cite{ferdman:cloudsuite} for our studies. In order to
capture the full-system interactions between the OS and these
workloads, we use a modified version of Pin~\cite{pin} - that reports
both virtual and physical addresses - to record 10-billion memory
traces containing virtual and physical pages, information about page
sizes allocated by the OS, references from kernel activity, and
information on page table modifications. Page table modifications are
tagged to identify situations when base pages are promoted to
superpages and vice-versa. We pass these traces to a carefully
designed and calibrated software simulation framework that models a
Sandybridge-style architecture with a detailed TLB and memory
hierarchy. We use an exhaustive set of studies on Cacti to model the
hardware with the correct timing and energy parameters.
Figure~\ref{fig:superpage-access_L1-hit-rate} shows the percentage of
memory references that fall on superpages across the workloads, and
the hit rates with a 32kB L1 cache.

\begin{center}
\small
\begin{tabular}{| l | c | c | c | c | c | c |} 
\hline
\multicolumn{3}{|c}{} & \multicolumn{4}{|c|}{\bf Access Latency (cycles)}\\ 
\hline
{\bf Cache} & {\bf VIPT} & {\bf Fre-} & {\bf TLB} & {\bf TLB} & {\bf L1 } & {\bf  L1 } \\ 
{\bf Size} & {\bf Assoc-} & {\bf quency} & {\bf base-} & {\bf super-} & {\bf base-} & {\bf super-} \\
{\bf (kB)} & {\bf iativity} & {\bf (GHz)} & {\bf page} & {\bf page} & {\bf page } & {\bf  page} \\ 
\hline
32  & 8 & 1.33	&	2		&	1	&			2	&		1 \\ 
\hline
32  & 8 & 2.80	&	4		&	2	&			4	&		2 \\ 
\hline
32  & 8 & 4.00	&	5		&	3	&			5	&		3 \\ 
\hline
64  & 16 & 1.33	&	5		&	1	&			5	&		1 \\ 
\hline
64  & 16 & 2.80	&	9		&	2	&			9	&		2 \\ 
\hline
64  & 16 & 4.00	&	13		&	3	&			13	&		3\\ 
\hline
128 & 32 & 1.33	&	14		&	2	&			14	&		2\\ 
\hline
128 & 32 & 2.80	&	30		&	3	&			30	&		3\\ 
\hline
128 & 32 & 4.00	&	42		&	4	&			42	&		4\\ 
\hline
\end{tabular}
\captionof{table}{\bf \small L1 Cache Configurations}
\label{table:config}
\end{center}

\begin{figure}[tb]
\setlength{\abovecaptionskip}{0pt}
\setlength{\belowcaptionskip}{0pt}
\begin{center}
\includegraphics[width=0.95\linewidth]{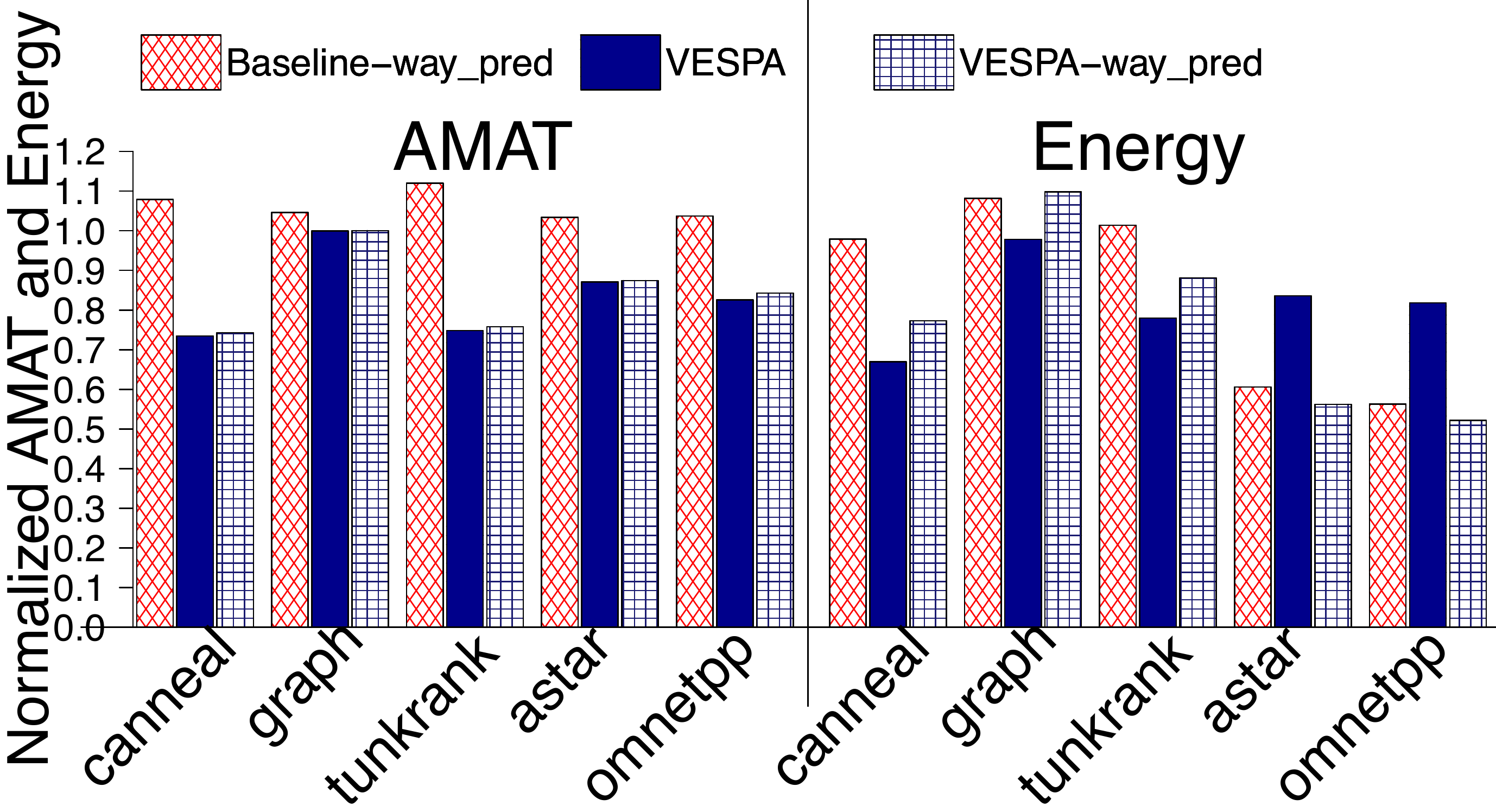}
\caption{\small Comparison of way-prediction with VESPA architecture
on 64kB size cache. Graph shows that benefits from way-prediction are
tied to its accuracy and depend on workload unlike VESPA, which
consistently gives lower Energy and AMAT than baseline.}
\vspace{-4mm}
\label{fig:waypred}
\end{center}
\end{figure}

\begin{figure*}[tb]
\setlength{\abovecaptionskip}{0pt}
\setlength{\belowcaptionskip}{0pt}
\centering
\subfloat[\small Superpage Allocation Probabilty = 25\%]
{
\includegraphics[width=0.33\linewidth]{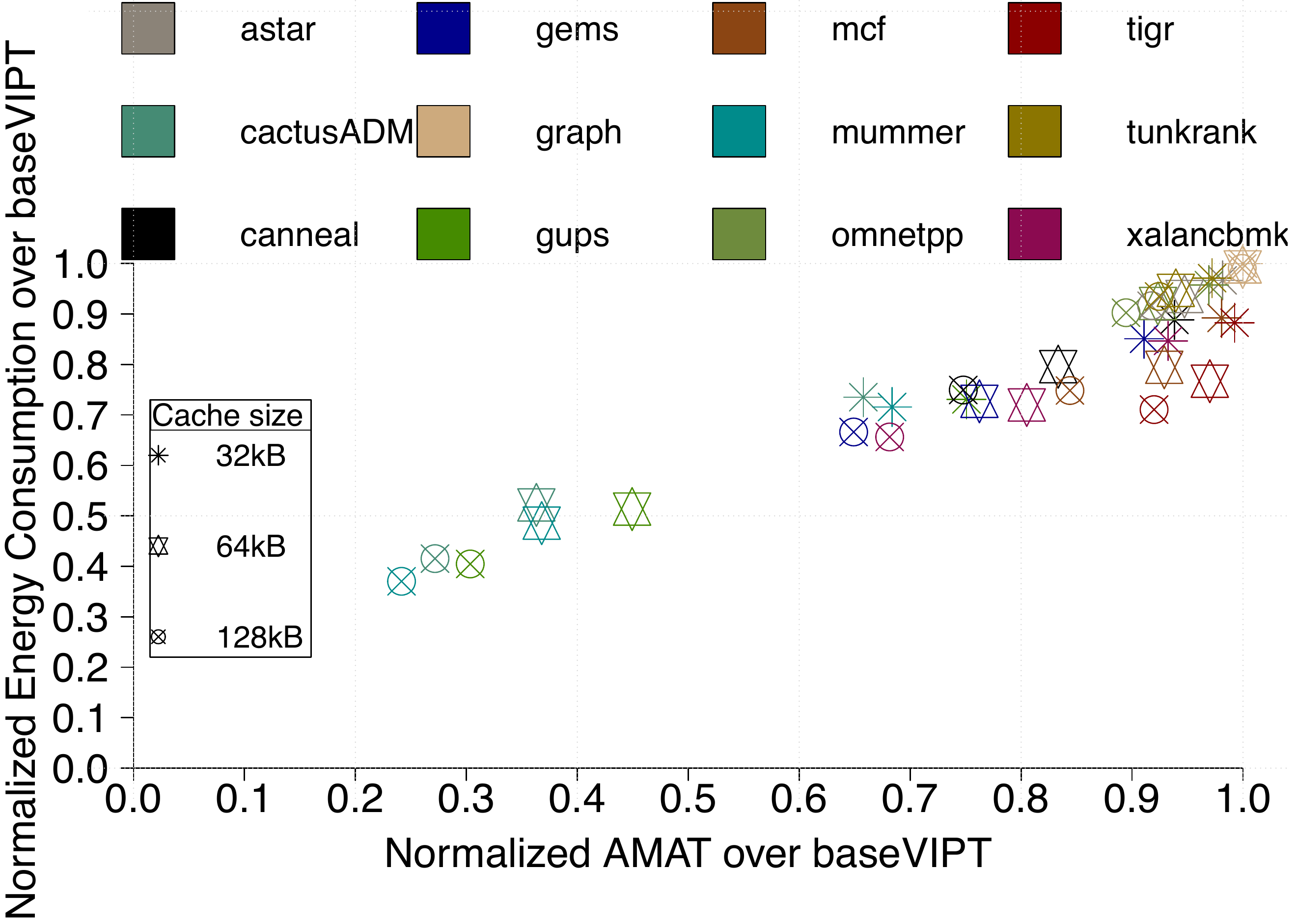}
\label{fig:25_frag}
}
\subfloat[\small Superpage Allocation Probabilty = 50\%]
{
\includegraphics[width=0.33\linewidth]{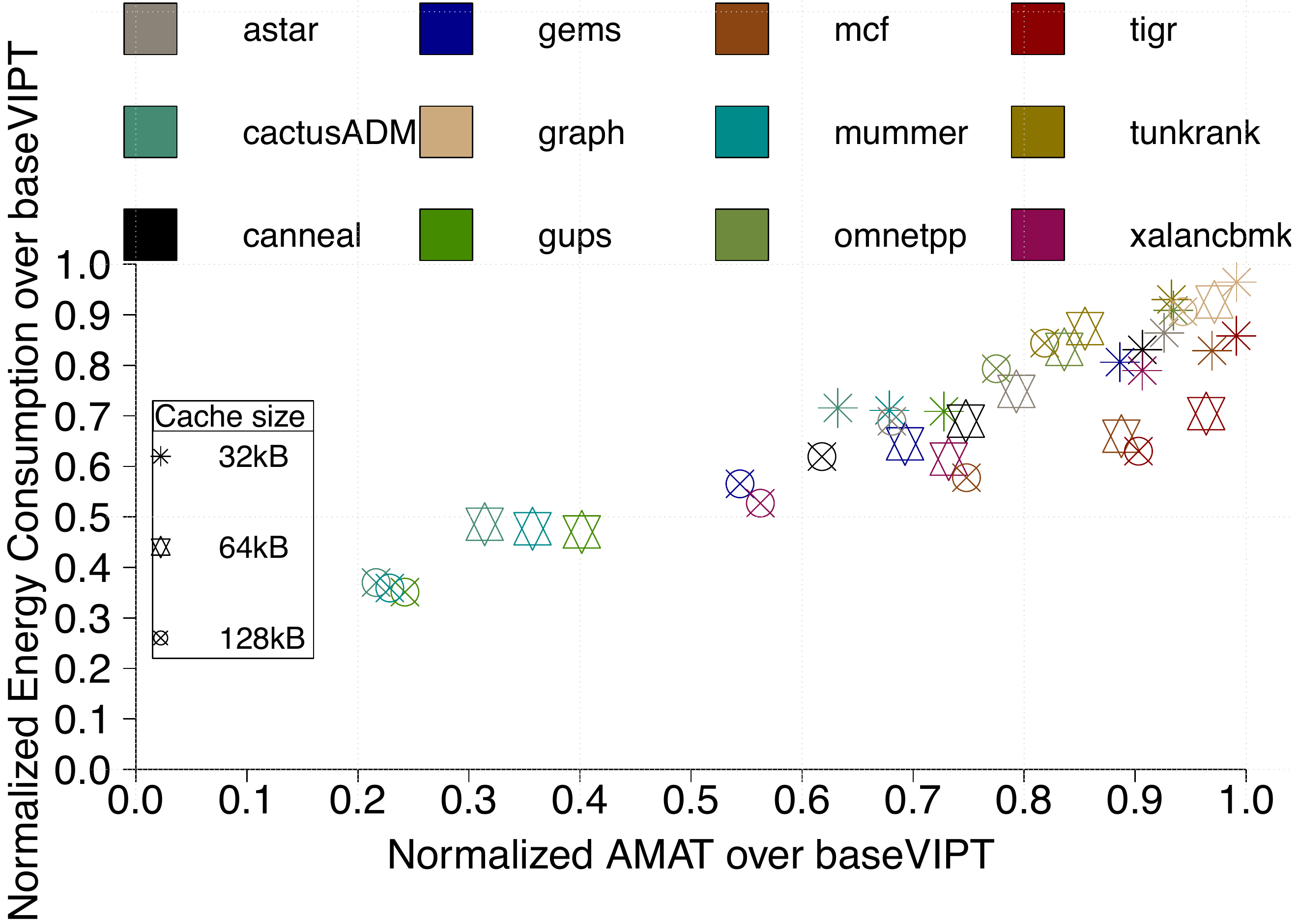}
\label{fig:50_frag}
}
\subfloat[\small Superpage Allocation Probabilty = 75\%]
{
\includegraphics[width=0.33\linewidth]{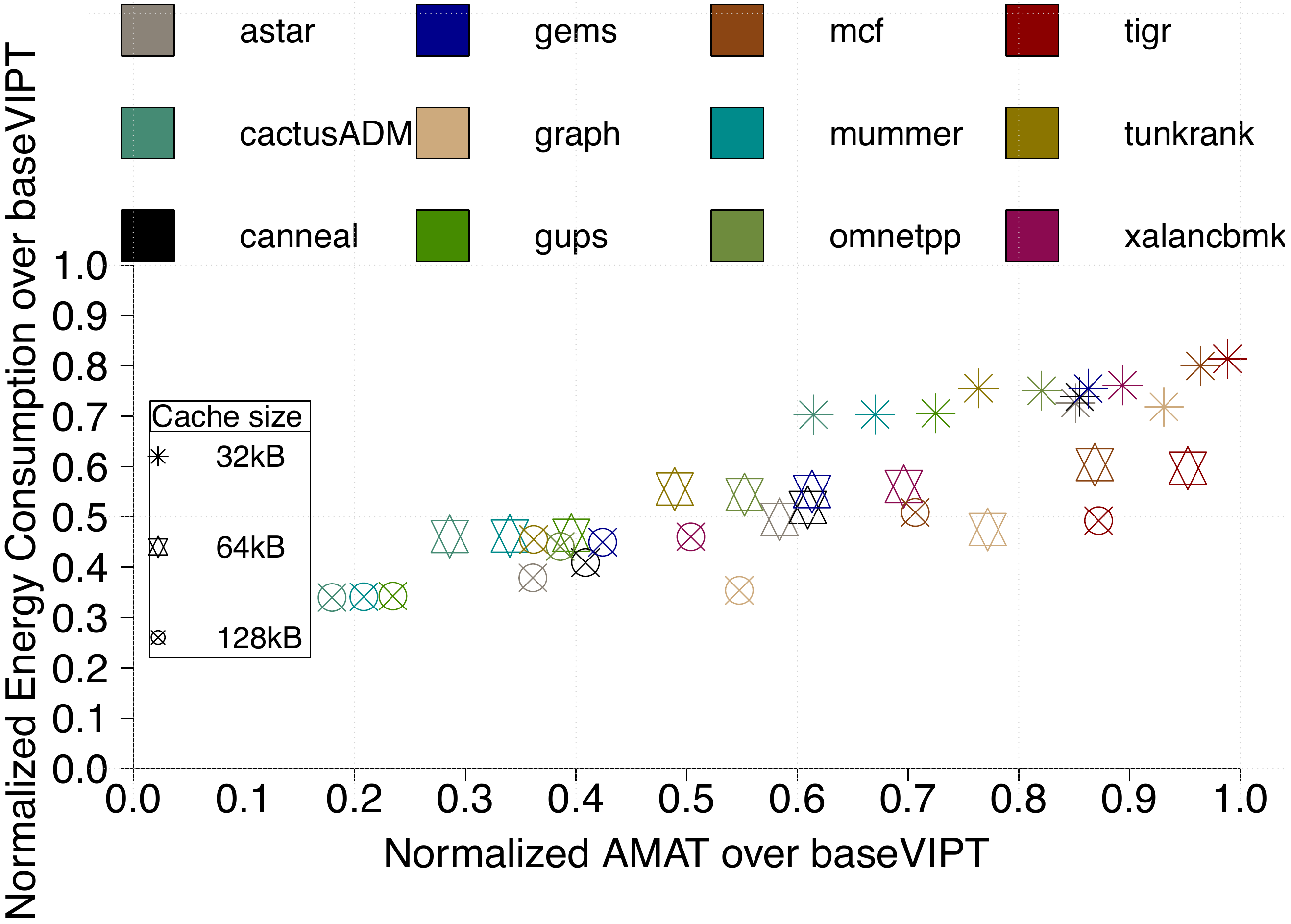}
\label{fig:75_frag}
}
\caption{\small \bf AMAT and energy reduction with VESPA as a function of memory fragmentation. The x-axis represent normalized AMAT for an
application with respect to their corresponding BaseVIPT cache, similarly y-axis represent
normalized energy with respect to corresponding BaseVIPT cache for a given application.}
\vspace{-4mm}
\label{fig:fragmentation_sweep}
\end{figure*}

\subsubsection{Energy}
\label{sec:eval_energy}
Figure \ref{fig:leakage_DynamicEnrgyWrkld} shows the normalized
improvement in total L1 access energy for 32kB, 64kB and 128kB L1
caches respectively, across the workloads.  Energy benefits are seen
across all the workloads, even for those that showed low AMAT
reductions (such as {\sf tigr}).  This is because of two reasons.
First, for both L1 hits and misses for superpages, {\sf VESPA} saves
dynamic energy, as Section~\ref{sec:anatomy} discussed.  Second,
leakage energy is saved per application as its overall runtime
decreases.  On average we see 8.92\% and 77\% reduction in dynamic and
leakage energy respectively with 32kB VESPA, 17.81\% and 95\%
reduction in dynamic and leakage energy respectively with 64kB VESPA
and 22.24\% and 98.89\% reduction in dynamic and leakage energy with
128kB VESPA over their respective baseline VIPT L1 caches.

\subsubsection{Performance}
\label{sec:eval_performance}
Figure \ref{fig:amat_wrkld} shows {\sf VESPA's} normalized AMAT
improvement across their respective baseline VIPT caches, for bigger
caches.  That is, {\sf VESPA-32kB} shows the improvement over {\sf
  BaseVIPT-32kB}, {\sf VESPA-64kB} shows the improvement over {\sf
  BaseVIPT-64kB}, and so on.  Some workloads such as {\sf canneal} and
{\sf gems} show up to 10-40\% reduction in AMAT.  This is because 60\%
of their accesses fall on superpages as
Figure~\ref{fig:superpage-access_L1-hit-rate} shows.  Others such as
{\sf cactusADM, gups} and {\sf mummer} show negligible reduction as
less than 10\% of their accesses are to superpages.  Workloads such as
{\sf tigr} exhibit interesting behavior. Even though 70\% of the
references are to superpages, AMAT reduces by only 1\% at 32KB, and
marginally increases with larger caches.  The reason is a low L1 hit
rate of 48\% as Figure~\ref{fig:superpage-access_L1-hit-rate} shows.
{\sf VESPA} is an optimization for the L1 cache; if the workload's
working set does not fit in L1, or if workload is streaming in nature,
leading to high L1 misses, then the AMAT is dominated by L2/DRAM
access latencies which are an order of magnitude higher than L1 access
latency and are not benefited by {\sf VESPA} in terms of AMAT.  But
even in these cases, {\sf VESPA} still saves energy as we show in
Section~\ref{sec:eval_energy}.  On average, {\sf VESPA} provides
4.45\% improvement in 32kB {\sf VESPA}, 12.09\% improvement in 64kB
{\sf VESPA} and 18.44\% improvement in 128kB {\sf VESPA} in AMAT, over
their respective baselines.


\begin{figure*}[!t]
\setlength{\abovecaptionskip}{0pt}
\setlength{\belowcaptionskip}{0pt}
\centering
\subfloat[\small Normalized improvement in AMAT in a 32-core system as the probability of superpage allocation increases.]
{
\includegraphics[width=0.28\linewidth]{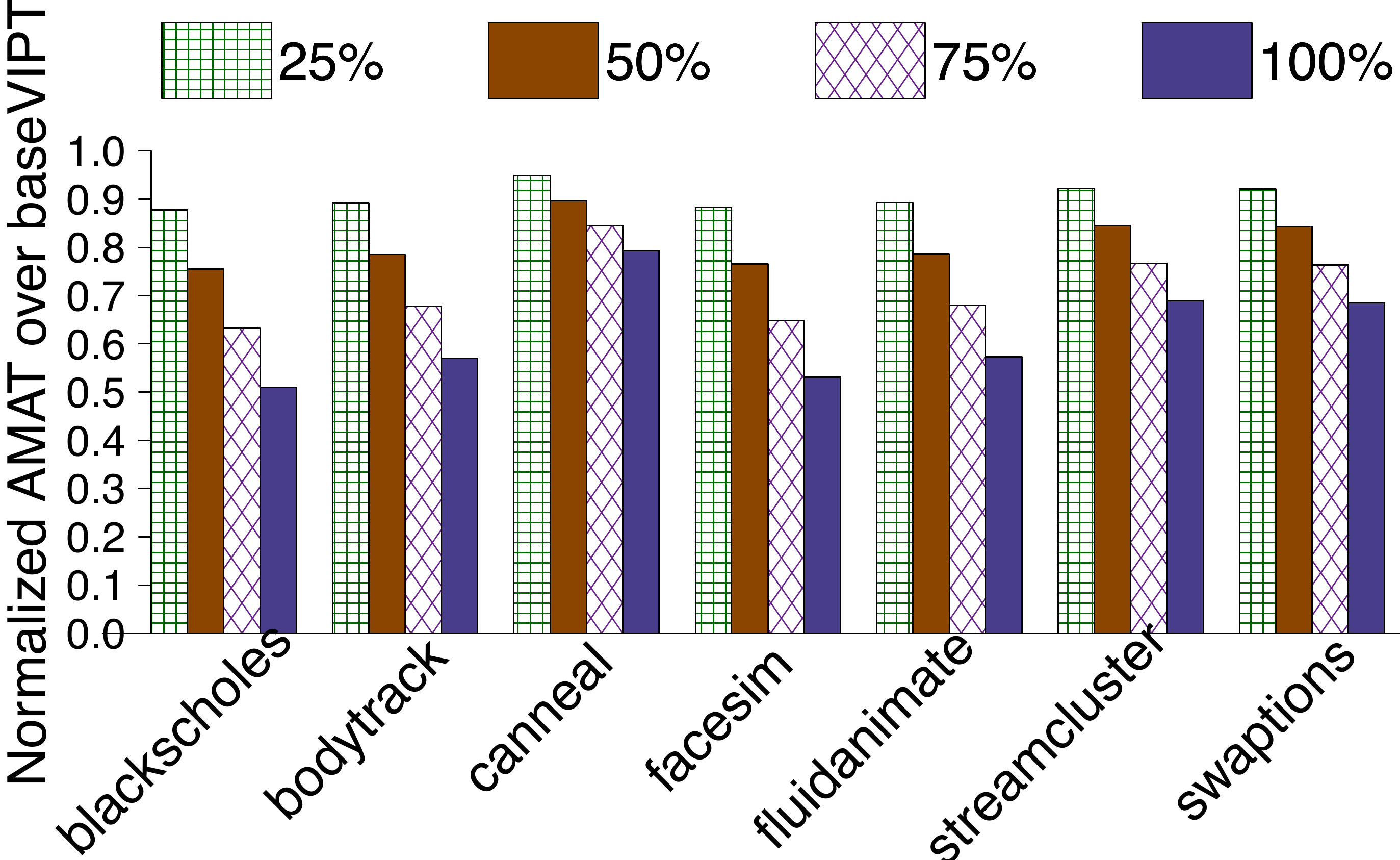}
\label{fig:parsec_amat}
}
\subfloat[\small  Normalized improvement in Dynamic Energy in a 32-core system as the probability of superpage allocation increases.]
{
\includegraphics[width=0.28\linewidth]{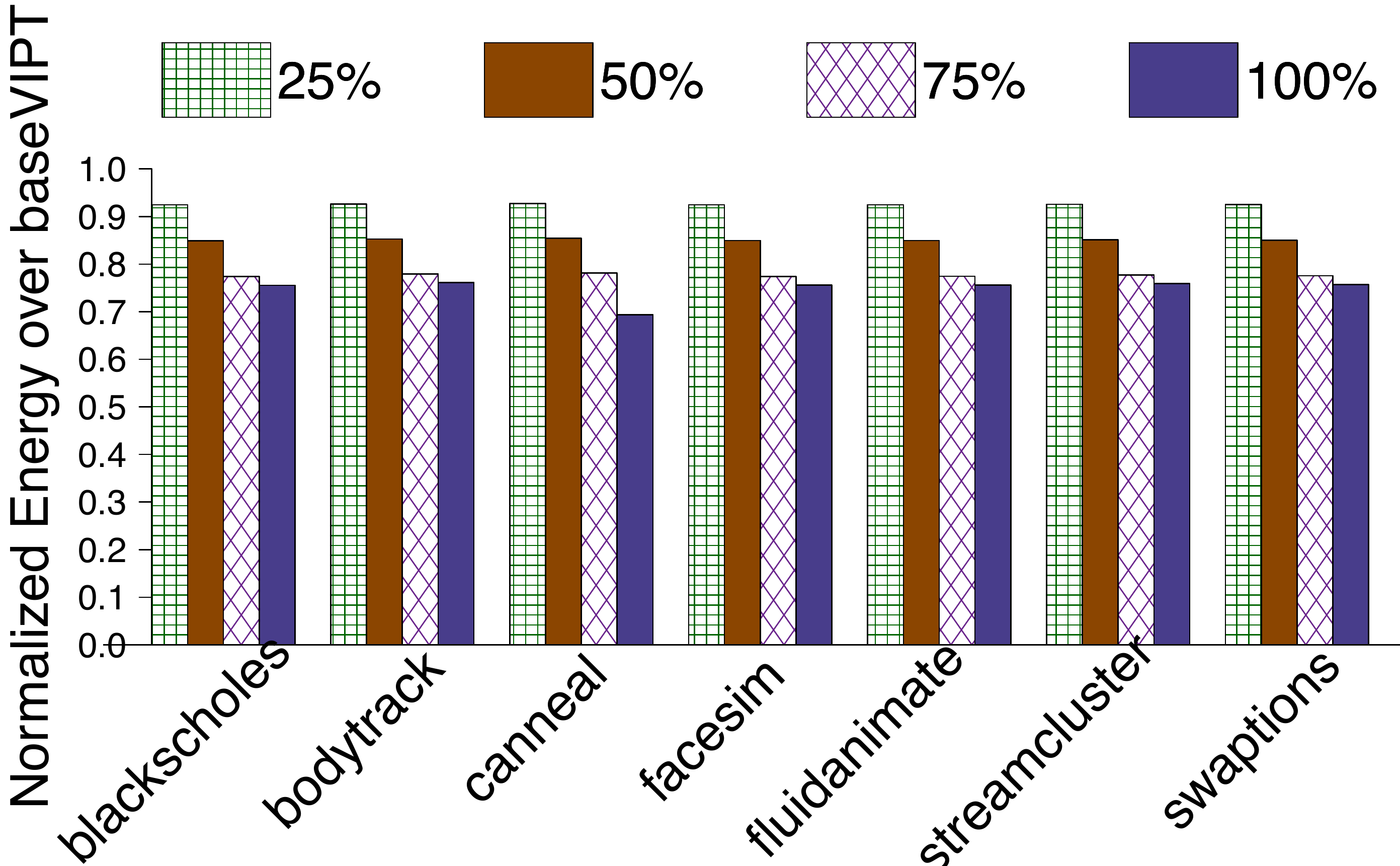}
\label{fig:parsec_energy}
}
\subfloat[\small Energy consumption for coherence lookups in L1, as  core counts 
increase, normalized to the 16-core BL energy for each benchmark.]
{
\includegraphics[height=3.7cm]{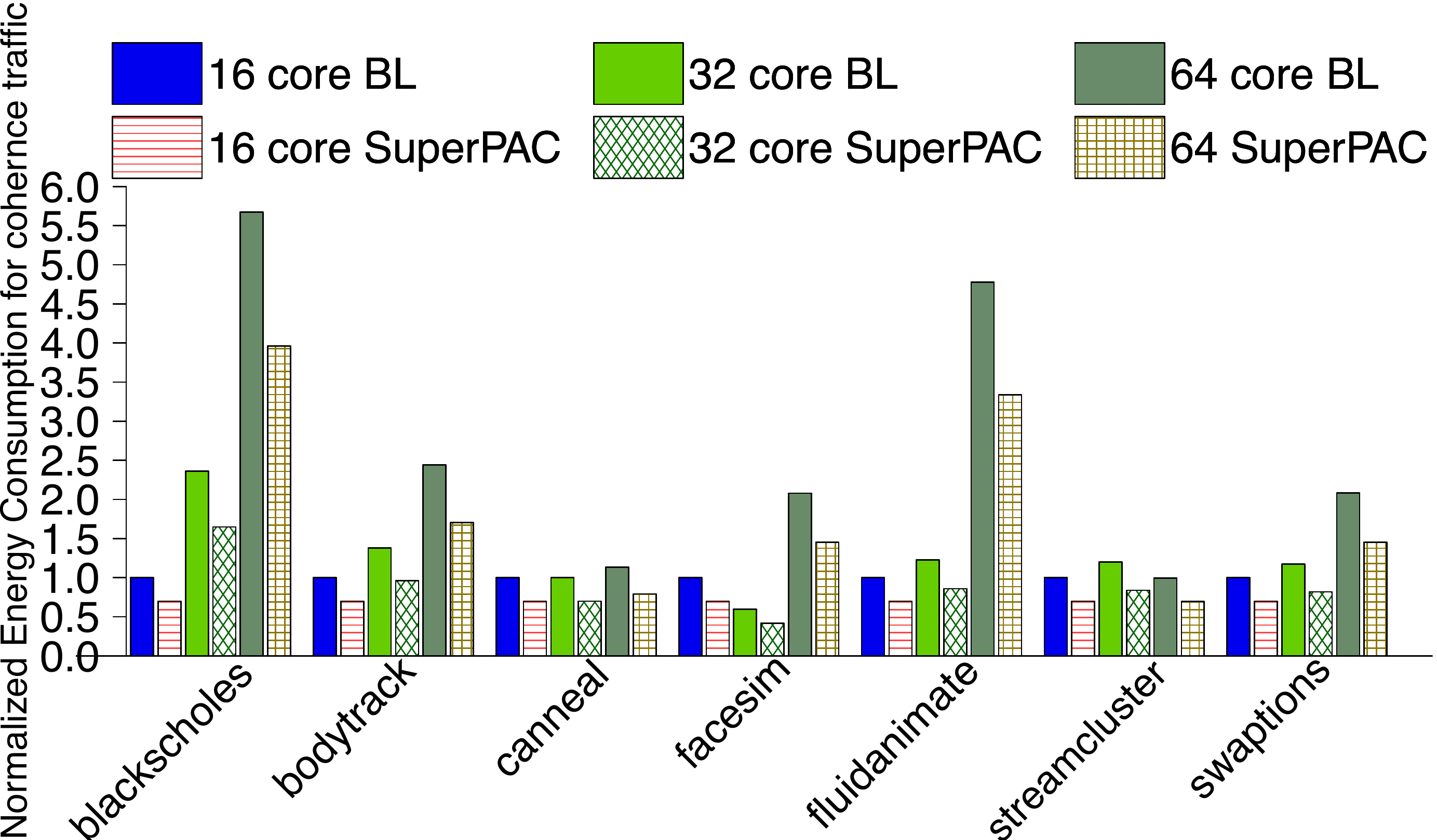}
\label{fig:parsec_coh_energy}
}
\caption{\small \bf Performance and Energy benefits of {\sf VESPA} in a Multicore System.}
\label{fig:parsec}
\end{figure*}

\subsubsection{Impact of Way Prediction}
\label{sec:eval_way-prediction}

Figure \ref{fig:waypred} quantifies the characteristics of {\sf VESPA}
versus a design with way-prediction. On the left, we plot AMAT values
while on the right, we plot energy values. All values are normalized
to a design with neither {\sf VESPA} no way-prediction. Note that
lower AMAT and energy values are desirable. We plot data from three
separate designs -- a design with just way-prediction (using an MRU
predictor as per prior work \cite{powell:reducing}), just {\sf VESPA},
and a combination of {\sf VESPA} and way-prediction. Figure
\ref{fig:waypred} reveals the following.

First, standard way-prediction always degrades AMAT. This is expected
since way-prediction trades access latency for better
performance. When prediction accuracy is good (e.g., for {\sf astar}
and {\sf omnetpp} which have prediction accuracy over 75\%), AMAT goes
up only marginally. But when MRU prediction suffers because workloads
use pointer-chasing memory access patterns with poorer access locality
(e.g., {\sf canneal}, {\sf graph500}, and {\sf tunkrank}), way
prediction can increase AMAT significantly. In contrast, {\sf VESPA}
can never degrade performance. At worst, it maintains baseline
performance in the absence of superpages. Far more commonly (as Figure
\ref{fig:waypred} shows), AMAT is improved dramatically.

Second, way-prediction can improve energy. However, in cases when
superpages are ample (e.g., for {\sf canneal}, {\sf graph500}, and
{\sf tunkrank}), {\sf VESPA} saves even more energy. In other cases,
when way-prediction is effective, {\sf VESPA} actually saves even more
energy when applied atop way-prediction (see the {\sf VESPA-waypred}
results). Therefore, in every single case, {\sf VESPA} remains
beneficial and orthogonal. Further, Figure \ref{fig:waypred} reveals
the potential symbiosis between {\sf VESPA} and way-prediction. We
intend studying more advanced schemes that dynamically choose when to
combine {\sf VESPA} and way-prediction, in future work.

\subsubsection{Effect of Memory Fragmentation}
\label{sec:fragmentation}

We perform a sensitivity study on how energy and latency of access gets affected
as the percentage of memory covered by superpages changes, by running applications 
along with {\sf Memhog}  in the background,
which was described earlier in Section~\ref{sec:os_support}.
We define {\it superpage allocation probability} as the probability of a page being a superpage, 
which goes down as memory fragmentation increases..

In Figure~\ref{fig:fragmentation_sweep} we plot normalized AMAT and L1
access energy reduction with varying cache sizes as a scatter plot, as
the superpage probability increases.  For workloads with high L1 hit
rates (Figure~\ref{fig:superpage-access_L1-hit-rate}), such as {\sf
  cactusADM, gups} and {\sf mummer}, we see the points moving towards
the origin as the superpage probability increases, demonstrating {\it
  both} AMAT and energy reductions.  For those with low hit rates,
such as {\sf tigr, mcf} and {\sf graph}, the AMAT reduction remains
low but the plots move vertically down as superpage probability
increases, demonstrating increased energy reduction.  For 32kB caches,
at a 75\% probability of superpages, we see up to 17.17\% reduction in
AMAT and 25.6\% reduction in L1 access energy for a 32kB cache. The
improvements go up to 56.72\% and 58.56\% respectively for 128kB.

\subsection{Multicore results}
\label{sec:eval_multicore}

\subsubsection{Methodology and workloads}
To observe the effect of {\sf VESPA} in multicore systems, 
we performed full-system simulations in gem5 \cite{gem5} for 
16, 32, and 64-core systems, running a directory-based 
MOESI protocol.
gem5's x86 model boots Linux v2.6
that does not have support 
for superpages without the {\sf hugetlbfs} filesystem, as modern Linux does. 
We mimic superpage support by adding our own shadow page table inside the memory system 
that maps every virtual address being sent to the memory system either on a base page 
or on a superpage, based 
on a superpage allocation probability (Section~\ref{sec:fragmentation}).
We run the PARSEC2.0 benchmark suite~\cite{parsec}, and study the AMAT and energy savings at the L1 
for both demand lookups from the core, as well as coherence lookups from the L2/directory.
All evaluations use a 32KB Private L1.

\subsubsection{Performance and energy for demand lookups}
Figure \ref{fig:parsec_amat} shows the normalized improvement in AMAT
for demand lookups for a 32 core system as a function of the superpage
allocation probability.  We see an AMAT reduction of 9.46\% on average
at 25\% probability, which goes up to 37.85\% at 100\% probability.
Figure \ref{fig:parsec_energy} similarly shows the normalized
improvement in dynamic energy consumed by the L1 for the system for
the same configuration.  We see a dynamic energy reduction of 7.45\%
on average at 25\% probability, increasing up to 25.16\% at 100\%
fragmentation.

\subsubsection{Energy savings for coherence lookups}
As discussed in Section \ref{sec:VESPA line insertion}, {\sf VESPA}
can reduce energy for {\it all} coherence traffic coming to L1,
irrespective of whether it is for data on a base page or a superpage,
since the right bank can be looked up from the physical address.  We
studied the savings in energy for coherence lookups (i.e., remote
loads, remote stores, invalidates, and writeback requests) at the L1
as the number of cores go up.  Figure \ref{fig:parsec_coh_energy}
plots our observations.  Data is normalized to the coherence energy
consumption of a 16 core {\sf BaseVIPT} configuration.  As the number
of cores go up, the total energy consumed by coherence lookups also
goes up as expected since the number of L1's has gone up. For
benchmarks with heavy sharing, such as {\sf blackscholes} and {\sf
  fluidanimate}, coherence lookup energy goes up by 5-5.5$\times$
going from 16-core to 64-core.  For others like {\sf canneal} and {\sf
  streamcluster}, it remains fairly flat.  {\sf VESPA} reduces dynamic
energy for each lookup by 30\%, with higher absolute energy savings as
core counts go up.  Note that these savings are for a full-bit
directory protocol that only send coherence lookups to the actual
sharers. Scalable commercial protocols, such as AMD
HyperTransport~\cite{ht}, that use limited-directories that
occasionally broadcast would show even more benefits with {\sf VESPA}.

\section{Related Work}\label{sec:related}
The challenges of VIPT caches have been an area of active study for
several years. Prior work has proposed VIVT caches \cite{energy_eff,
  Segmented_addr, VirCache, kaxiras2013new, SynonymFilter} as an
alternative, obviating the need for TLB lookup before L1 cache
access. While VIVT caches are attractive because they decouple the TLB
and L1 cache, they remain hard to implement because of the challenges
of maintaining virtual page synonyms, the difficulties of correctly
managing multiple processes and their context switches, and their
interactions with standard cache coherence protocols which operate on
physical addresses. While recent work does present interesting and
effective solutions to the problems of synonyms \cite{VirCache,
  SynonymFilter} and cache coherence \cite{kaxiras2013new}, they
require non-trivial modifications to L1 cache and datapath design. At
the same time, work on {\it opportunistic virtual caching} \cite{ovc}
proposes an L1 cache design that caches some lines with virtual
addresses, and others (belonging to synonym virtual pages) as physical
addresses. And finally, as an alternative to hardware enhancements,
past work has proposed modifying OS page allocation techniques to
prevent synonym problems on virtually-addressed caches
\cite{mayank-please-fill}.

Unlike prior work, we minimally modify the L1 cache to increase the
flexibility of VIPT, not replace it entirely. As a result, unlike
prior work, we do not require sophisticated prediction logic,
significant changes to coherence protocols, or changes to the OS or
applications stack. We exploit already-existing OS optimizations
(i.e., superpages) and repurpose them to attack the difficulties of
traditional VIPT.

\section{Conclusion}
L1 caches are critical for system performance as they service every
cacheable memory access from the CPU and coherence lookups from the
underlying memory hierarchy.  Their design involves a delicate balance
between fast lookups, low access energy, high hit rates, and
simplicity of implementation.  In this work, we identify the
opportunity presented by superpages in virtual memory systems today,
to optimize current VIPT L1 caches. Our design, {\sf VESPA}, provides
performance improvements, and energy (dynamic + leakage) reduction for
all L1 lookups - both CPU initiated and coherence initiated.  We add
modest hardware changes, and no changes to the page table or the OS.
We believe that {\sf VESPA} will become even more crucial in future as
L1 cache sizes increase to handle larger working sets of big data
applications.  To the best of our knowledge, this is the first work to
optimize L1 caches for superpages.
\bibliographystyle{ieee}
\bibliography{ref}

\end{document}